\documentclass[conference]{IEEEtran}
\IEEEoverridecommandlockouts

\makeatletter
\long\def\@makecaption#1#2{\ifx\@captype\@IEEEtablestring%
\footnotesize\begin{center}{\normalfont\footnotesize #1}\\
{\normalfont\footnotesize\scshape #2}\end{center}%
\@IEEEtablecaptionsepspace
\else
\@IEEEfigurecaptionsepspace
\setbox\@tempboxa\hbox{\normalfont\footnotesize {#1.}~~ #2}%
\ifdim \wd\@tempboxa >\hsize%
\setbox\@tempboxa\hbox{\normalfont\footnotesize {#1.}~~ }%
\parbox[t]{\hsize}{\normalfont\footnotesize \noindent\unhbox\@tempboxa#2}%
\else
\hbox to\hsize{\normalfont\footnotesize\hfil\box\@tempboxa\hfil}\fi\fi}
\makeatother

\usepackage{cite}
\usepackage{epsfig}
\usepackage{subfigure}
\usepackage{amsfonts} 
\usepackage{bbold}
\usepackage{amsmath}

\usepackage{mathtools}          
\usepackage{color}
\ifCLASSINFOpdf

\newcommand{\Vijay}[1]{\BREAKME{\color{red} [#1]}}

\begin{document}

\title{
Adaptive Semi-Persistent Scheduling for Enhanced On-road Safety in Decentralized V2X Networks
}


\author{Avik Dayal$^{1}$\thanks{This research is supported by the Office of Naval Research (ONR) under MURI Grant N00014-19-1-2621. This paper has been accepted for publication in IFIP Networking 2021.
This is a preprint version of the accepted paper.}, Vijay K. Shah$^{1}$,  Biplav Choudhury$^{1}$, Vuk Marojevic$^{2}$, Carl Dietrich$^{1}$, and Jeffrey H. Reed$^{1}$ \\
$^{1}$Wireless@VT, Bradley Department of ECE, Virginia Tech, VA, USA \\
$^{2}$Electrical and Computer Engineering, Mississippi State University, MS, USA \\
\{ad6db, vijays, biplavc, cdietrich, reedjh\}@vt.edu and vuk.marojevic@ece.msu.edu   }


\date{June 2020}

\maketitle


\begin{abstract}

Decentralized vehicle-to-everything (V2X) networks (i.e., Mode-4 C-V2X and Mode 2a NR-V2X), rely on periodic Basic Safety Messages (BSMs) to disseminate time-sensitive information (e.g., vehicle position) and has the potential to improve on-road safety. For BSM scheduling, decentralized V2X networks utilize sensing-based semi-persistent scheduling (SPS), where vehicles sense radio resources and select suitable resources for BSM transmissions at prespecified periodic intervals termed as Resource Reservation Interval (RRI). In this paper, we show that such a BSM scheduling (with a fixed RRI) suffers from severe under- and over- utilization of radio resources under varying vehicle traffic scenarios; which severely compromises timely dissemination of BSMs, which in turn leads to increased collision risks. To address this, we extend SPS to accommodate an adaptive RRI, termed as SPS++. Specifically, SPS++ allows each vehicle -- (i) to dynamically adjust RRI based on the channel resource availability (by accounting for various vehicle traffic scenarios), and then, (ii) select suitable transmission opportunities for timely BSM transmissions at the chosen RRI. Our experiments based on Mode-4 C-V2X standard implemented using the ns-3 simulator show that SPS++ outperforms SPS by at least $50\%$ in terms of improved on-road safety performance, in all considered simulation scenarios.


\end{abstract}

\begin {IEEEkeywords}
Decentralized V2X, C-V2X, NR-V2X, Semi-Persistent Scheduling, Basic Safety Message, On-road Safety
\end{IEEEkeywords}

\section{Introduction}
\label{intro}
Vehicle-to-everything (V2X)~\footnote{V2X refers to vehicle-to-infrastructure (V2I), vehicle-to-vehicle (V2V), and vehicle-to-pedestrian (V2P) communications.} communications is a promising technology for next generation of intelligent transportation systems (ITS), mainly due to its potential of improving on-road safety, leading to the prevention/reduction of road accidents and more efficient traffic management \cite{V2V_safety}. There are two competing technologies that enable V2X communications, Dedicated Short Range Communications (DSRC) and Cellular-V2X (C-V2X) \cite{magazine_paper}. Dedicated Short Range Communications (DSRC) is a decentralized wireless technology based on the 802.11p standard\cite{TTC_IEEE_ACCESS}.
On the other hand, 3GPP introduced C-V2X communications in Release 14, and standardized two modes of operation termed Mode 3 and Mode 4 based on the scheduling preferences. Mode 3 C-V2X employs a centralized scheduling 
approach under the coverage of eNodeB, where two vehicles can communicate directly. The selection of radio resources are managed by the control signaling from the cellular infrastructure over the Uu interface (uplink and downlink) \cite{electronics8080846}. Mode 4 C-V2X adopts new PC5 interface for direct communication among vehicles without the need for coverage from the eNodeB \cite{CV2X_Congestion}. Since the cellular connectivity can not be assumed ubiquitous, Mode 4 C-V2X is considered as the baseline mode for C-V2X. Recently, New Radio V2X (NR-V2X) has been envisioned in Release 16, and it is expected to have decentralized Mode 2a NR-V2X \cite{Naik_survey_5G}.


 
There are two major V2X use cases: (i) \textit{Cooperative safety applications} that rely on \textit{basic safety messages (BSMs)}, which are periodic messages that contain critical safety information, e.g., sender vehicle's position and speed, and (ii) \textit{Cooperative traffic efficiency messages}, which are event triggered messages that are intended to help vehicular flow. 
Since BSMs carry time-sensitive information, BSMs enable cooperative safety applications, such as, forward collision warnings \cite{sengupta2007cooperative} and blind spot/lane change warnings \cite{Huang_IEEENetwork}. This is mainly because BSMs facilitate \textit{accurate positioning} or \textit{localization} of neighboring vehicles. Outdated BSMs (due to large BSM scheduling intervals) and/or lost BSMs (due to channel congestion) negatively impacts the performance of safety applications as it leads to increased collision risky situations, mainly due to wrong localization of neighboring vehicles, or in other words, \textit{high tracking error}. Tracking error can be defined as the difference between a vehicle's actual and perceived location (via most recent BSM) by its neighboring vehicles. 
From the above discussion, it is evident that BSM scheduling (or time intervals at which BSMs are broadcasted) becomes a critical parameter for ensuring BSM timeliness and minimal channel congestion, and thus, improved on-road safety of vehicles. 

In Mode 4 C-V2X, the BSM scheduling is achieved by utilizing sensing-based semi-persistent scheduling (SPS), where the vehicle sense the radio resources (or channel medium) and select suitable (unutilized or underutilized) radio resources for the transmission of BSMs at prespecified fixed time intervals, termed, \textit{Resource Reservation Interval} (RRI). More formally, RRI can be defined as the inter-transmission time interval between two consecutive BSM transmissions by a vehicle. The value of RRI is usually set to $100$ ms (equivalently, BSM rate to $10$ Hz) in the C-V2X standard. The 3GPP standard allows other values of RRI, such as, $20$ ms and $50$ ms \cite{magazine_paper}. 

In this paper, we show that the SPS algorithm with fixed RRIs (e.g., $100$ ms) has limitations in terms of improving on-road safety performance of Mode-4 C-V2X networks. Consider the following instances: (i) \textit{high vehicle density scenarios} -- C-V2X networks would likely suffer from overly congested radio channels (and thus, large number of lost or dropped BSM packets), which will lead to increased tracking error and thus, degraded on-road safety, and (ii) \textit{low vehicle density} -- The radio resource are under-utilized in such low density scenarios. The on-road safety can be greatly improved by choosing lower value of RRIs (e.g., $20$ ms), as lower value of RRIs will improve the timeliness of BSMs without compromising the channel congestion (as the channels can support lower RRIs for fewer vehicles). Furthermore, notice that due to vehicle mobility and other contextual factors, the vehicle densities and available channel resources may change over time even for a given C-V2X scenario. 
Motivated by this, we propose significant enhancements to conventional (fixed RRI) SPS, and develop an improved, adaptive BSM scheduling algorithm, termed SPS++. At any time slot, SPS++ allows each vehicle to adapt (or adjust) the RRI for a given network scenario (depending upon vehicle density around it), and thus, ensure timely BSM transmissions, which in turn, promises to improve on-road safety performance. Decentralized Mode 2a NR-V2X is expected to utilize SPS as the standard BSM scheduling protocol, thus, we anticipate that our SPS++ algorithm will also be applicable to the new NR-V2X standard\cite{Naik_survey_5G}.


In summary, the paper makes following key contributions:

\begin{itemize}
\item We show that the conventional SPS algorithm for Mode-4 C-V2X suffers severely from under- and over-provisioning of radio channel resources (depending upon the vehicle densities). This in turn negatively impacts the timely successful delivery of BSMs and thus, compromises on-road safety performance of Mode-4 C-V2X. 

\item To address the limitations of SPS, we propose an improved, adaptive SPS, termed SPS++, that ensures BSM timeliness and thus, improves on-road safety performance of Mode-4 C-V2X. SPS++ allows each vehicle to dynamically adjust RRIs at each time slot, and judiciously utilize radio resources for BSM transmissions while accounting for varying C-V2X vehicle traffic scenarios.

\item To measure the on-road safety performance of C-V2X, we present collision risk model, which is based on the concepts of tracking error and time-to-collision (TTC).

\item Our extensive experiments based on ns-3 simulator show that the proposed SPS++ greatly outperforms the conventional SPS in terms of road safety performance, measured as collision risk, across all considered C-V2X vehicle scenarios. When compared to SPS with RRI as $20$ ms, $50$ ms, and $100$ ms, the results show a significant improvement in the reduction of collision risk respectively by $51.27$\%, $51.20$\%, and $75.41\%$ for $80$ vehicles/km. 

\end{itemize}

The rest of the paper is organized as follows: Section \ref{Section 2 Related Work} discusses the related works. Section \ref{section 3- System Model} discusses Mode-4 C-V2X and SPS algorithm. Section \ref{Section 4:On Road Safety Performance} discusses on-road safety performance model. In Section \ref{Section 5- SPS++}, we highlight the limitations of SPS in the context of on-road safety performance, and propose SPS++ algorithm. Section \ref{Section- Simulation Description and Results} describes the performance evaluation, followed by concluding remarks in Section \ref{Section-Conclusion}.

\section{Related work}
\label{Section 2 Related Work}
Recent works have simulated and proposed resource allocation improvements to Mode-4 C-V2X. Molina-Masegos et al. \cite{magazine_paper} and Chen et al. \cite{IOTJ_2016_V2X_Chen} develop and use simulations to assess the performance of the SPS algorithm in highway and city scenarios, and showed the improved performance of C-V2X over DSRC. Research in \cite{Nabil_SPS}, \cite{Bazzi_CV2X}, and \cite{CV2X_Congestion} investigate the performance of parameters used in SPS such as RRI, probability of reselection, and selection window on the packet delivery ratio (PDR). Gonzalez-Martin et al. \cite{Analytical_CV2X} build the first known analytical model of SPS based C-V2X. Recently, \cite{IOTJ_SPS} proposed short term sensing before resource selection to help reduce packet collisions and improve SPS performance. Halder et al. \cite{electronics8080846} and Lee et. al \cite{Lee_TENCON_2020} suggest adjusting the transmission power and RRI respectively, to improve the overall performance of SPS, but both only focus on improving the PDR performance of C-V2X.

As opposed to prior works which focus on communication-centric metrics such as, throughput and PDR, we focus on on-road safety performance of decentralized V2X networks, such as, tracking error and collision risks.
It is worth mentioning that designing optimal beacon rate in case of DSRC for improved safety performance of DSRC based V2X networks has been explored fairly well~\cite{TTC_IEEE_ACCESS, choudhury2020experimental, kaul2011minimizing_AOI, Dayal_VTC,ETSI_COOP_AWARENESS_2015}. However, note that designing rate control algorithms for DSRC are fundamentally different than designing scheduling protocols for decentralized Mode-4 C-V2X networks, and existing rate control protocols can not be directly applied to improving on-road safety performance in our context of Mode-4 C-V2X networks. \textit{To the best of our understanding, this is the first work which investigates designing BSM scheduling protocol for improved on-road safety performance of Mode-4 C-V2X.}

\section{C-V2X Mode-4}
\label{section 3- System Model}
\par In this section we discuss the Mode-4 C-V2X standard based on 3GPP Release 14. We start with the C-V2X Mode 4 physical layer and then detail the inner workings of sensing-based SPS that allows vehicles to find and reserve suitable transmission opportunities. 

\subsection{Physical Layer}
\par At the physical layer, Mode-4 C-V2X is similar to the LTE uplink and uses single-carrier frequency division multiple access (SC-FDMA). C-V2X supports 10 and 20 MHz channels at 5.9 GHz. In Mode-4, the time-frequency resources are divided into resource blocks (RBs), subframes ($sf$), and subchannels. Frames are 10 ms in length, and are comprised of 10 subframes. Subframes are typically comprised of $2$ resource blocks in time \cite{Bazzi_CV2X}. A resource block is the smallest schedulable unit of time and frequency that can be allocated to users. Each RB is made up of 7 SC-FDMA symbols and is 180 kHz wide in frequency with 12 equally spaced subcarriers. Each RB occupies 0.5 ms in time, with a minimum allocation time interval of 1 ms, also referred to as the transmission time interval (TTI). Subchannels are composed of resource blocks contiguously located in frequency. In C-V2X, the data packet to be transmitted, called the Transport Block (TB), typically takes up one or more subchannels. Each TB is transmitted in the same subframe along a sidelink control information (SCI), that contains the modulation and coding scheme (MCS) \cite{CV2X_Congestion}. 



\subsection{Semi-Persistent Scheduling (SPS)}
\label{Section- SPS Background}
\begin{figure}[h]
\centering
\includegraphics[width=\columnwidth]{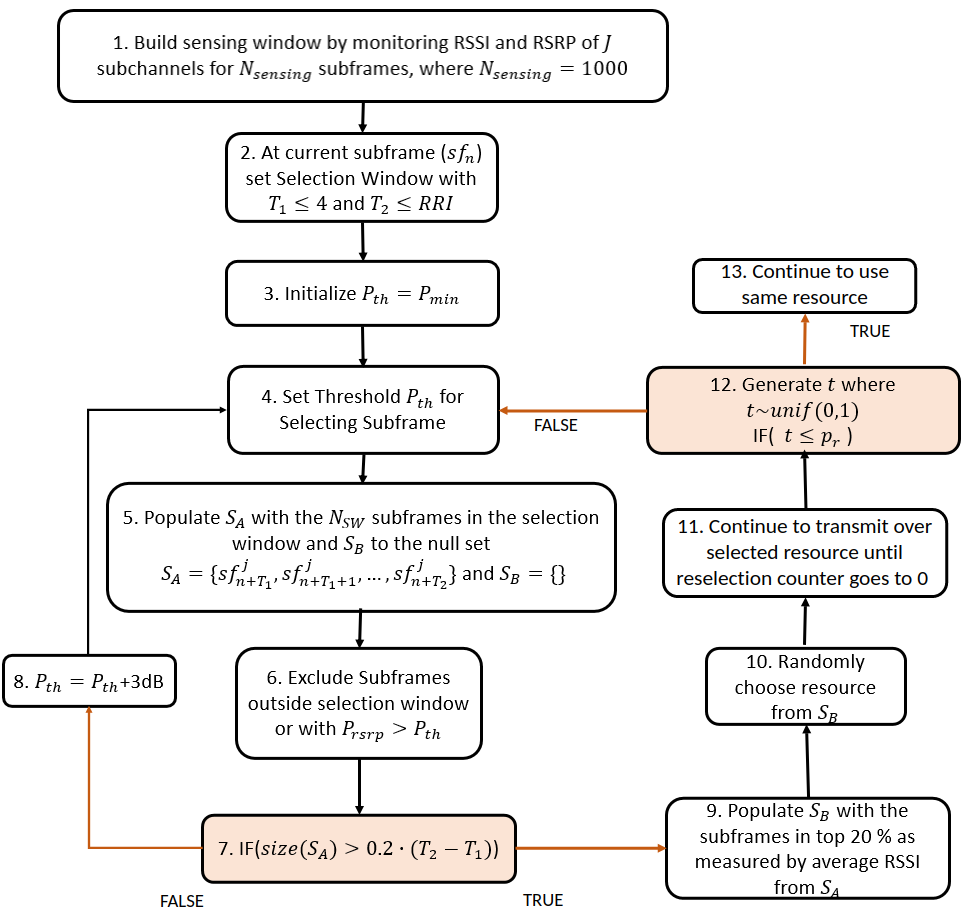}
\caption{Flowchart of SPS Algorithm }
\label{SPS_algorithm_fig}
\centering
\end{figure}

\begin{figure*}[!h]
\subfigure[
\label{SPS_illustration}]{
\epsfig{figure=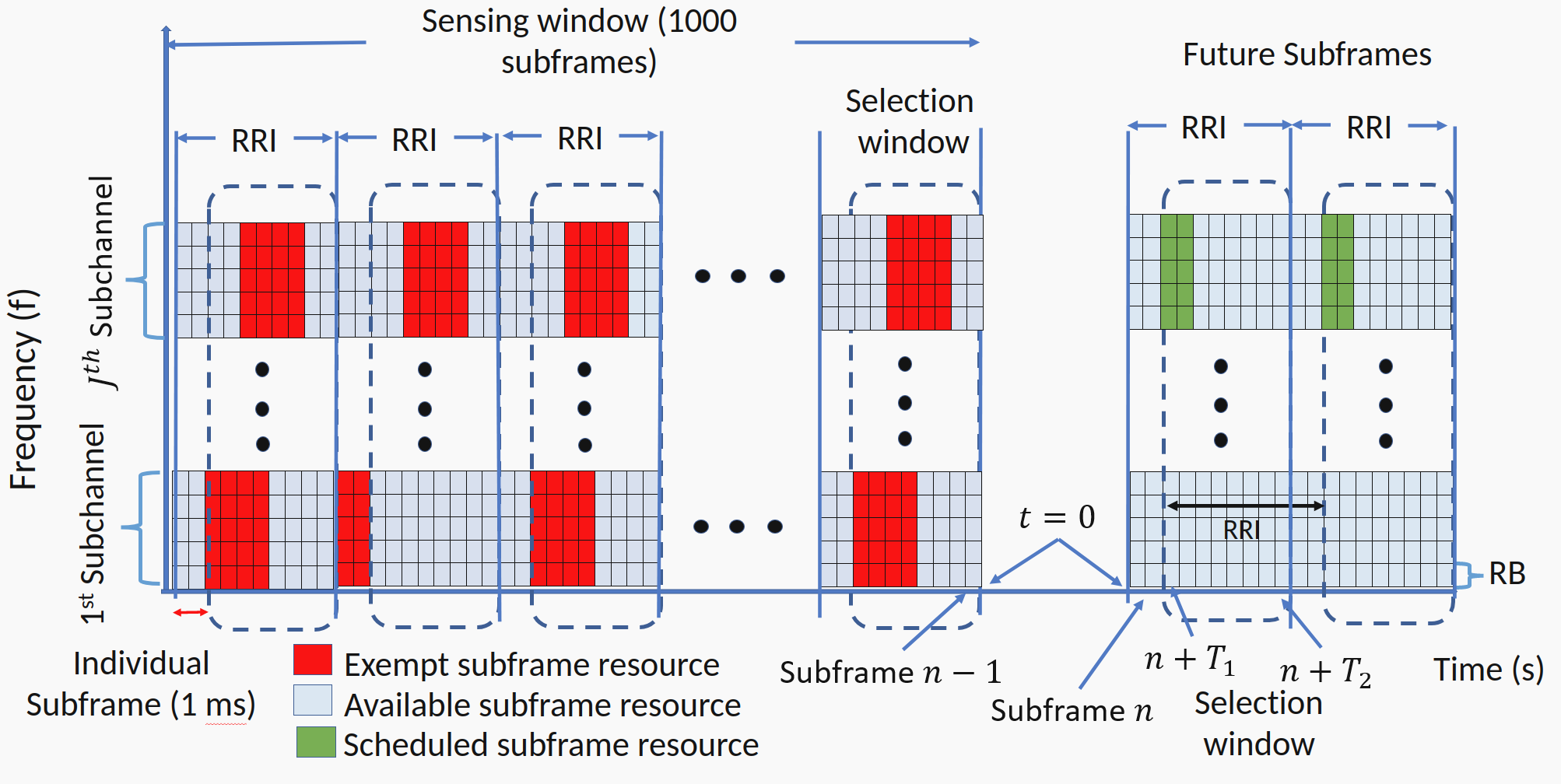,width=4.25 in,  keepaspectratio}}
\subfigure[\label{tracking_error_fig}]{
\epsfig{figure=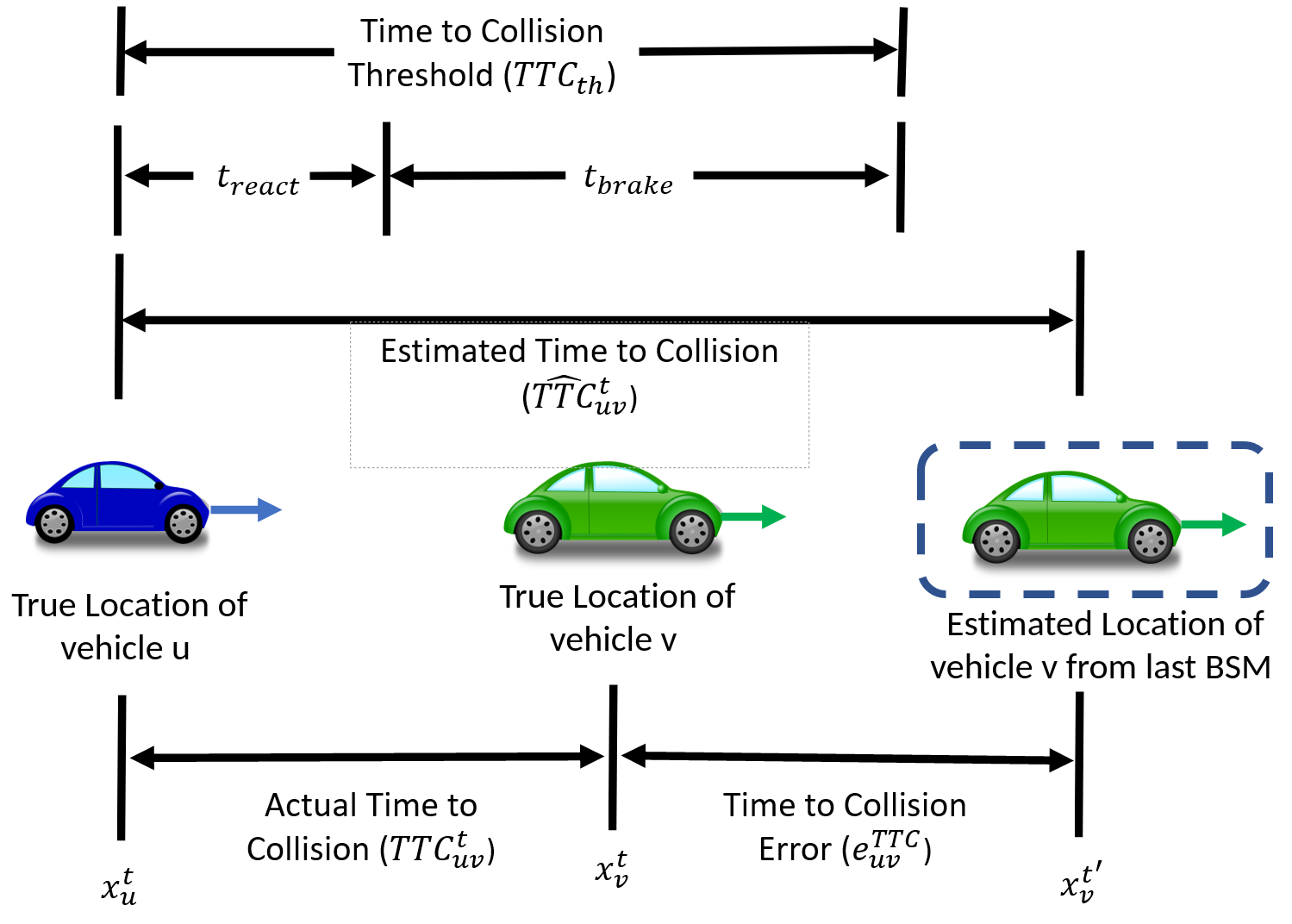,width=2.72 in,  keepaspectratio}}
\caption{(a) Illustration of the sensing window, selection window, RRI, and scheduling for SPS and (b) Illustration of Collision Risky scenario caused by a large tracking error.}\vspace{-0.2in}
\end{figure*}


At the MAC layer, Mode-4 C-V2X utilizes SPS that uses sensing to determine suitable semi-persistent transmission opportunities, i.e., set of subframes, for BSM transmission. Fig. \ref{SPS_algorithm_fig} depicts the SPS algorithm~\footnote{Please refer to \cite{magazine_paper} and \cite{Bazzi_CV2X} for a detailed discussion on SPS.}, and is explained below. We use $sf_{i}^{j}$ to refer to a single-subframe resource where $i$ is the subframe index and $j$ is the subchannel index of $J$ total subchannels. See Fig. \ref{SPS_illustration} for the illustration of subframe and other key terminologies (e.g., sensing window) related to SPS.

\begin{itemize}
    
    \item \textbf{Sensing (Step 1):} Each vehicle continuously monitors the subframes by measuring the reference signal received power (RSRP) and the sidelink received signal strength indicator (S-RSSI) across all $J$ subchannels; and stores sensing measurements for a prespecified last $N_{sensing}$ subframes, known as the \textit{sensing window}. $N_{sensing}$ is set to $1000$ subframes. Let subframe $sf_n$ denote the first subframe after the sensing window. Then we can write the sensing window at $sf_n$ as the following set of single-subframe resources for the $j^{th}$ subchannel: $\left[sf_{n-N_{sensing}}^{j}, \dots, sf_{n-1}^{j}\right]$.
    
    
    \item \textbf{Identifying available resources (Steps 2-8):} 
    Each vehicle initializes a \textit{selection window} with a set of consecutive candidate subframes (See Step 2). $T_1 \leq 4$ and $T_2 \leq $ \textit{RRI} are the start and end subframes for the selection window. 
    The \textbf{RRI} refers to the time interval between two consecutive BSM transmissions (See Fig. \ref{SPS_illustration} for illustration). Each vehicle utilizes $N_{sensing}$ subframes (obtained in Step 1) for identifying and subsequently, selecting the available subframe  within the selection window for BSM transmission
    as follows. 
    \begin{enumerate}
        \item The vehicle sets -- (i) the RSRP threshold, $P_{th}$, to a minimum RSRP value, $P_{min}$ (Step 3) and (ii) initializes set $S_A$ as all subframes in the selection window, i.e., $S_A=[sf_{n+T_1}, sf_{n+T_1+1}, \cdots, sf_{n+T_2}]$, and $S_B$ as an empty set (See Step 5). 
        \item As shown in Step 6, the vehicle excludes all candidate subframes from set $S_A$ if one of the following conditions are met -- (i) the vehicle has not monitored the corresponding candidate subframe in the sensing window (i.e., $N_{sensing}$) and (ii) the linear average RSRP measurement for corresponding candidate subframe is higher than $P_{th}$. The RSRP exclusion criteria for the $i^{th}$ subframe (for $j$th sub-channel) in the selection window can be written as 
        
        \begin{equation} \label{RSRP_calculation1}
            \hspace*{-1.5cm}     \frac{1}{\mathcal{K}}\sum_{k=0}^{\mathcal{K}}RSRP\left(sf_{n+T_1+i-N_{sensing}+k \cdot RRI}^{j}\right)\geq P_{th}
        \end{equation}
        where $\mathcal{K} = \frac{N_{sensing}}{RRI}$. 
        If $RRI=100$ ms and $N_{sensing} = 1000$ ms and $i = 4$, then, we look at RSRP value across following $10$ subframes -- $\{4, 104, 204, \dots, 904\}$ and divide it by $\mathcal{K} = \frac{1000}{100} = 10$ to find the average RSRP.
        
        \item If the remaining subframes in $S_A$ is less than $20\%$ of total available subframes (Step 7), then $P_{th}$ is increased by $3$ dB (Step 8), and Steps $4$ to $7$ are repeated.
    \end{enumerate}

    \item \textbf{Resource Selection (Steps 9-10):} 
    If more than $20\%$ of available channel resources are identified, then, as shown in Step 9, the vehicle populates $S_B$ with the first $20\%$ of candidate subframes which has the lowest average S-RSSI in set $R_A$. The vehicle then randomly selects a candidate subframe from set $S_B$ as selected resource for first BSM transmission (See Step 10).
    \begin{equation} \label{RSSI_calculation1}
            RSSI\left(\sum_{k=0}^{\mathcal{K}}sf_{n+T_1+i-N_{sensing}+k \cdot RRI}^{j}\right)
    \end{equation}
    
    
    \item \textbf{Resource Reselection (Steps 11-13):} Each vehicle can reserve the same subframe (selected in Step 10) for next \textit{Resource Counter (RC)~\footnote{Resource Counter (RC) is the maximum number of transmissions a certain vehicle is allowed (by utilizing the selected subframe/resource in the current selection window) before having to reselect a new set of resources.}} number of subsequent transmissions with the same transmission interval, i.e., \textit{RRI}. 
    The RC varies with the RRI to ensure that the selected subframe/resource is in use for at least 0.5 s and at most 1.5 s. This means that for a 20 ms RRI, $25\leq \textnormal{RC} \leq 75$, for 50 ms RRI, $10\leq \textnormal{RC} \leq 30$, and for a 100 ms RRI, $5\leq \textnormal{RC} \leq 15$
    
    \hspace{0.2in}After RC reaches $0$, the vehicle can either continue utilizing the preselected resources with a probability $p_r$ or reselect new resources for BSM transmissions with a probability $(1 - p_r)$ (See Steps 11-13). 
    
\end{itemize}

\section{On-Road Safety Performance}
\label{Section 4:On Road Safety Performance}
In this section, we propose a collision risk model, which measures the on-road safety performance of Mode-4 C-V2X networks. Our proposed collision risk model is inspired from risk model presented in literature \cite{choudhury2020experimental}, and is based on tracking error and time-to-collision (TTC) as discussed below.


\subsection{Tracking Error (TE)}
\label{Subsection:Tracking Error}
Tracking error (TE), $e^{track}_{uv}$, is defined as the difference between the ground truth location of the sender vehicle $u$ and $u$'s location as estimated by the neighboring receiver vehicle $v$. Let the most recent BSM received at $v$ from $u$ was generated at time $t^{'}$. At time $t > t'$, the tracking error (TE) that $v$ has in tracking $u$ can be calculated as follows: 


\begin{equation}\label{TE_eq}
    e^{track}_{uv} = |x^t_u - x^{t'}_u|
\end{equation}

$x^t_u$ is the actual location of $u$ at time $t$ and $x^{t'}_u$ is the $u$'s location information contained in the most recent BSM received from $u$. We consider $x$-coordinate to calculate the tracking error as the vehicle moves in x-direction only~\footnote{For the ease of presentation, we consider a simple tracking error (TE) model with no lateral movements across lane.}.
The TE is usually significant because

\begin{enumerate}
    \item Each BSM takes non-zero channel delay (e.g., propagation delay) to be successfully delivered to the receiver after being generated, and the sender would have moved a non-zero distance during the time period. 
    \item The RRI (or in other words, inter-BSM transmission interval) is significant. For example, if RRI is $100$ ms, it means there is at least $100$ ms inter-reception delay between two consecutive BSM from sender vehicle $u$ at the receiver $v$ even if the channel delay is zero. Thus, TE at vehicle $v$ would be significant as vehicle $u$ would have moved to a different location during this time interval.
    
    \item Lost or delayed BSM, mainly due to channel congestion, will further deteriorate TE.
\end{enumerate}

Note that lower TE value at receiver vehicle $v$ means that $v$ is able to track $u$ well (i.e., accurately position vehicle $u$). Thus, TE across all vehicles is used to measure TTC and Collision risk as discussed in next subsection.

\subsection{Time-to-Collision (TTC) and Collision Risk}
\label{subsection:TTC_Collision_risk}


TTC for a pair of vehicles is defined as the time needed for the distance between the two vehicles to become zero, which denotes a potential collision between them. We relate the collision risk (or on-road safety performance) to the TTC, which in turn, utilizes tracking error. 


At any time $t$, the receiver $v$ can estimate TTC with respect to its neighboring vehicle $u$ based on the BSM sent by $u$ as:

\begin{equation} \label{TTC_calculation1}
   \text{Estimated TTC}, \hspace{0.25cm} \widehat{TTC}^{t}_{uv} = \frac{|\widehat{x}^{t'}_u - x^t_v|}{s_{u,v}}
\end{equation}

where $\widehat{x}^t_u$ is $u$'s location as per the last received BSM at $v$ from $u$ with generation time $t'$ and $x^t_v$ is the v's actual location at time $t$. $s_{u,v}$ is the relative velocity between them. 

However, since sender $u$ is currently at location $x^t_u$ at time $t$ (and not $x^{t'}_u$), the true TTC at time t is:
\begin{equation} \label{TTC_calculation2}
   \text{True TTC}, \hspace{0.25cm} TTC^{t}_{uv} = \frac{|x^t_u - x^t_v|}{s_{u,v}}
\end{equation}

Using Equations \ref{TE_eq}, \ref{TTC_calculation1}, and \ref{TTC_calculation2}, we get


\begin{equation} \label{cum_TTC_eq}
    \widehat{TTC}^t_{uv} = TTC^t_{uv} + \frac{|e^{track}_{uv}|}{s_{uv}} \\
\end{equation}

We designate $\frac{|e^{track}_{uv}|}{s_{uv}}$ as the TTC error, $e^{TTC}_{uv}$. From Eq. \ref{cum_TTC_eq}, it is clear that the tracking error affects the TTC calculation. This estimated TTC, $\widehat{TTC}^t_{uv}$ can lead to collision risky situations, and its significance in collision warning was studied in \cite{shladover2006analysis}. For example, consider that $v$ overestimated/underestimated~\footnote{Both overestimating and underestimating TTC are hazardous as overestimating means the vehicle gets too close, and decision based on understimated values will impact the other vehicles.} the TTC to $u$ and is about to take a route/maneuver based on this errorneous value of $\widehat{TTC}^t_{uv}$. 
Such a situation can be safely avoided by manual intervention if the time taken by the driver to react and apply the brakes to make the vehicle stop, defined as, TTC threshold ($TTC_{th}$), is less than the true TTC (i.e., $TTC^t_{uv}$) at any given time $t$. Mathematically, TTC threshold, $TTC_{th}$ is given by:


\vspace{-0.2in}
\begin{eqnarray}
\label{TTC_threshold_eq}
    TTC_{th} = t_{brake} + t_{react}
\end{eqnarray}

where $t_{react}$ is 1s and represents the time taken by the driver to respond to the situation and apply the brakes\cite{brake_time_2000}. $t_{brake}$ is the time taken by the vehicle to complete to a stop after the brakes have been applied. Assuming vehicle $u$ has velocity $s_u$ and every vehicle has a maximum deceleration $a$ as 4.6 m/$s^2$ (from \cite{brake_time_2000}) makes $t_{brake}=s_u/a$. If the true TTC, i.e., $TTC^t_{uv}$, between a pair of vehicles exceeds the $TTC_{th}$ (i.e., driver's controllability), it results in a collision risky scenario.

It means, given $TTC_{th}$, and $TTC^t_{uv}$, the collision risk, $CR^t_{uv}$, for a given vehicle pair $u$ and $v$ at any given time $t$, can be computed as follows:

\begin{equation} \label{Collision_risk_eq}
    CR^t_{uv} = 
    \begin{cases}
        1 & TTC^t_{uv} \leq TTC_{th}\\
        0 & otherwise
    \end{cases}
\end{equation} 

Using Eq. \ref{cum_TTC_eq}, Eq. \ref{Collision_risk_eq} can be rewritten as:

\begin{equation} \label{Collision_risk_eq_2}
    CR^t_{uv} = 
    \begin{cases}
    
        1 & (\widehat{TTC}^t_{uv} - e^{TTC}_{uv})\leq TTC_{th}\\
        0 & otherwise
    \end{cases}
\end{equation} 

From Eq. \ref{Collision_risk_eq_2}, it is evident that improving on-road safety performance of C-V2X, i.e., reducing collision risky scenarios, is directly proportion to minimizing TE, $e^{track}_{uv}$, between each pair of vehicles in the C-V2X networks.

Fig. \ref{tracking_error_fig} illustrates how the TTC error causes a collision risky scenario. In Fig. \ref{tracking_error_fig}, vehicle $u$ estimates vehicle $v$ to have a TTC, $\widehat{TTC}^t_{uv}$ beyond $TTC_{th}$, i.e., in a "safe" position. However in reality, vehicle $u$ is close to vehicle $v$ and the true TTC $TTC_{uv}$ is within the $TTC^t_{th}$, and causes collision risky scenario. Thus, ideally TTC error should be non-existent (or zero) in order to prevent such a collision risky scenario. However, as discussed earlier, TTC error exists as the TE is significant (mainly, due to scheduling, transmission and communication delays). As per the recommendations by SAE-J2945~\cite{SAE_J2945}, there exists a TE threshold $e^{track}_{th} = 0.5$m which does not lead to a collision risky situation. We use this TE threshold to rewrite Collision risk in Eq \ref{Collision_risk_eq_2} as follows.
\begin{equation} \label{Collision_risk_eq2}
    CR^t_{uv} = 
    \begin{cases}
        1 & (e^{TTC}_{uv} > \frac{|e^{track}_{th}|}{s_{uv}}) \textbf{ and } ({TTC}^t_{uv} < TTC_{th}) \\
        0 & otherwise
    \end{cases}
\end{equation} 

$s_{u,v}$ is set to be the average relative velocity between a pair of vehicles $u$ and $v$. 

At any time $t$, using Eq. \ref{Collision_risk_eq_2}, we count the number of instances between each pair of vehicles in which TTC error exceeds the ratio $\frac{|e^{track}_{th}|}{s_{uv}}$ and true TTC, $TTC^t_{uv}$, is within TTC threshold, $TTC_{th}$, as the measure of collision risks. 

\section{Adaptive SPS++ Scheduling Protocol}
\label{Section 5- SPS++}

In this section, we present the limitations of conventional SPS protocol in terms of improving on-road safety performance of Mode-4 C-V2X, followed by detailed discussion on the proposed adaptive SPS++ scheduling protocol.

\subsection{Limitations of SPS on On-Road Safety Performance }

 \begin{figure}[h]
\centering
\includegraphics[width=0.75\columnwidth]{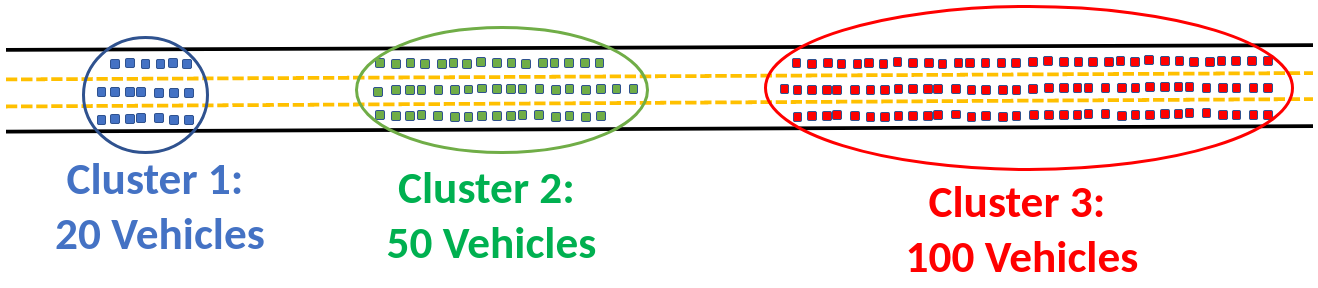}
\caption{C-V2X example network with three clusters of vehicles.}
\label{cluster_fig}
\centering
\end{figure} 

We present the limitations of SPS through a simple C-V2X example (See Fig. \ref{cluster_fig}). The example C-V2X network consists of three clusters~\footnote{Each vehicle is at 1-hop (i.e., within the transmission range) of every other vehicles belonging to a certain cluster of vehicle}, of vehicles, where cluster $1$, $2$, and $3$ respectively have $20$, $50$, and $100$ vehicles. We make the following \textbf{assumptions} for all example C-V2X scenarios.
\begin{itemize}
    \item We assume $T_1 = 0$ and $T_2 = $ RRI, which means, the size of selection window is equal to the RRI.
    \item C-V2X physical layer consists of $2$ subchannels only. Each BSM transmission uses both the subchannels and takes 1 ms to transmit. 
    This means if the selection window is $100$ ms (i.e., RRI = $100$ ms), then at most $100$ distinct vehicles have unique BSM transmission opportunities (assuming no collision in resource selection)

    \item Each cluster of vehicles is sufficiently spaced apart from each other so that there is no inter-cluster interference. This means, for example, no transmissions from cluster 2 interfere with any transmissions from cluster 1, and vice-versa.     
    
\end{itemize}

Under the above assumptions, let us look at the (i) \textit{Channel Occupancy Percentage $C_{occup}$}, defined as, the percentage of the number of vehicles transmitting to the total number of available subframes transmission opportunities., and (ii) \textit{Probability of Successful Reception ($P_{suc})$}. For simplicity, let $P_{suc}$ is given by $\frac{1}{N}$ where $N$ is the number of vehicles using the same subframe for BSM transmissions. (In reality, $P_s$ gets worse as the number of vehicles $N$ increases.) 


Table \ref{table:SPS_metrics} depicts the average $C_{occup}$ and $P_{suc}$ observed in the example C-V2X network under conventional SPS with three different values of RRIs, i.e., $20$ ms, $50$ ms, and $100$ ms. For instance, the average $C_{occup}$ for SPS with low RRI = $20$ ms is given by $\frac{\sum RRI \times (C_{occup} \text{in each cluster i})}{\text{total number of vehicles}} $. $C_{occup}$ for each cluster can be computed as follows:
Since RRI is $20$ ms, there are $20$ transmission opportunities (or subframes), (i) in cluster 1, $20$ vehicles attempt to transmit, it means $C^1_{occup} = \frac{20}{20} \times 100 = 100\%$, (ii) in cluster 2, $50$ vehicles attempt to transmit, which results in $C^2_{occup} = \frac{50}{20} \times 100 = 250\%$, and (iii) in cluster 3, $100$ vehicles attempt to transmit, resulting in $C^3_{occup} = 500\%$. Thus, the average $C_{occup}$ is $\frac{20 \times C^1_{occup} + 50 \times C^2_{occup} + C^3_{occup}}{20 + 50 + 100} \times 100  = 379.4\%$. The average $P_s$ can be computed in the similar fashion, and it turns out to be $0.35$ in case of SPS with RRI as $20$ ms. On contrary for SPS with high RRI = $100$ ms  the average $C_{occup}$ and $P_s$ are $75.88\%$ and $1$ respectively.

 \begin{table}[]
    \centering
     \caption{conventional SPS with fixed RRIs vs SPS++ scheduling with adaptive RRI}   
    \label{table:SPS_metrics}
    \begin{tabular}{|p{2.5cm}|p{0.75 cm}|p{0.75 cm}|p{0.75 cm}|p{1cm}|p{1.8cm}|}
    \hline
     Metrics & \multicolumn{3}{c}{Conventional SPS} & \textbf{SPS++}  \\ \hline
      & 20 ms & 50 ms & 100 ms & \textbf{Adaptive RRI}\\ \hline
     Channel Occupancy Percentage & 379.4\% & 152.94\% & 75.88\% & 100\% \\ \hline
     Probability of Successful Reception & 0.35 & 0.7058 & 1 & 1 \\ \hline
    \end{tabular}
    \vspace{-0.1in}
\end{table}
 
Note that SPS with low RRI such as, $20$ ms leads to \textit{overly congested radio channels} ($379.4\%)$, and thus, large number of dropped BSM packets ($0.35$), particularly, in clusters 2 and 3 with $>20$ vehicles). The lost packets result in high tracking error, which compromise the on-road safety performance of considered C-V2X network. Whereas, in case of SPS with high RRI as $100$ ms, the radio resources are under-utilized ($75\%$), particularly in cluster $1$ and $2$ with $<100$ vehicles. On-road safety performance can be significantly improved by choosing lower value of RRI as lower value of RRIs will improve timely delivery of BSMs. From the above discussion, it is evident that SPS with fixed RRI (irrespective of the chosen value of RRI) is limited in the context of improving overall on-road safety performance of C-V2X networks. 


To address the limitations of conventional SPS (and as detailed in next subsection), we propose an improved scheduling strategy, termed, SPS++, which allows each vehicle to adapt its RRI, based on its neighboring vehicle density at any given time. In case of example C-V2X network, under such SPS++ strategy, each vehicle in cluster 1 (with $20$ vehicles) will choose RRI = $20$ ms. Similarly, SPS++ will choose RRI = $50$ ms for cluster 2 (with $50$ vehicles) and RRI = $100$ ms for cluster 3 (with $100$ vehicles) -- which will result in $C_{occup} = 100\%$ and $P_{s} = 1$ (See Table \ref{table:SPS_metrics}). 
It means that the proposed SPS++ protocol strategy with adaptive RRI enables judicious utilization of the radio resources
This in turn reduces the tracking error and enhances the on-road safety of C-V2X networks.

\begin{figure}[h]
\centering
\includegraphics[width=\columnwidth]{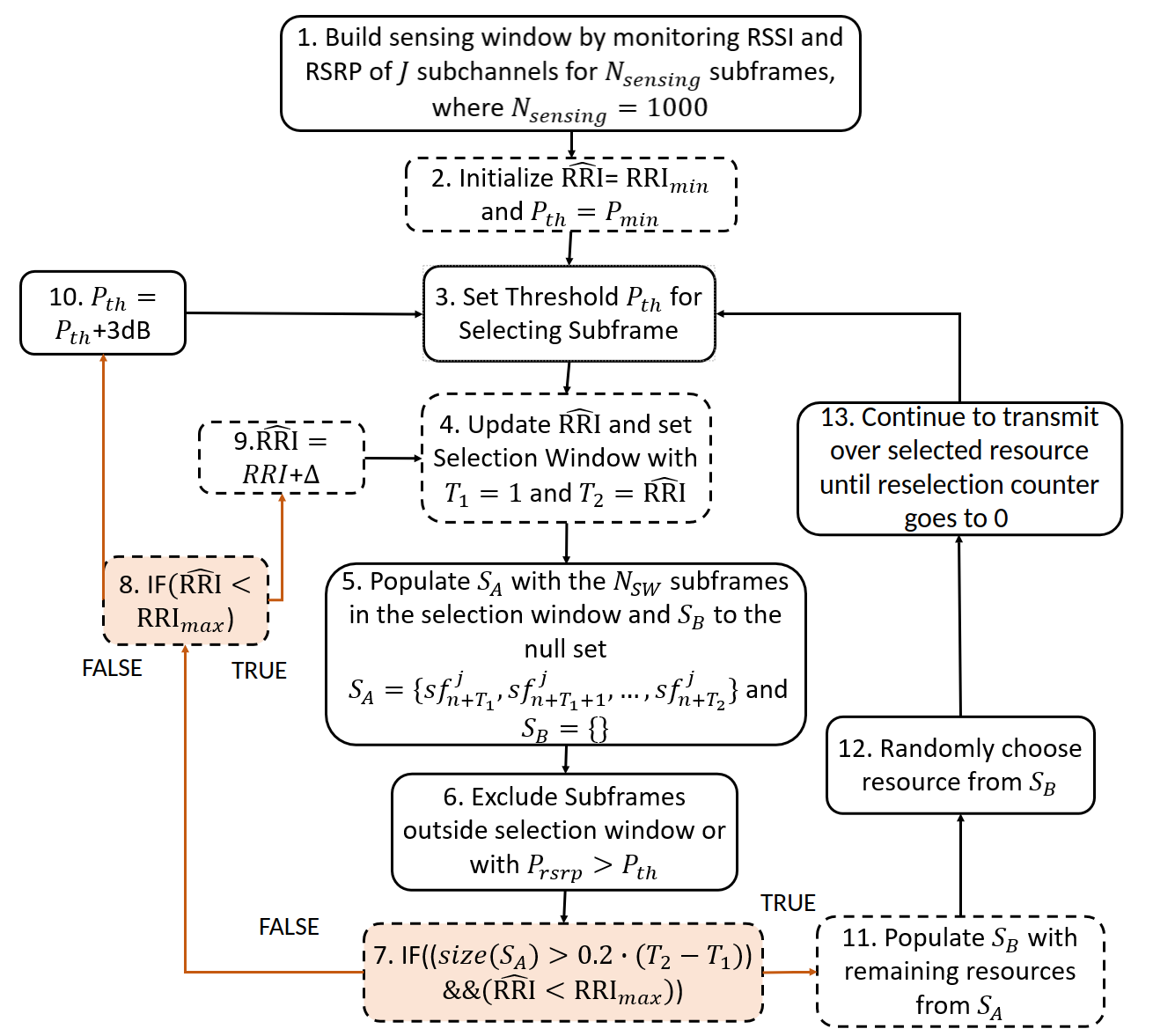}
\caption{Flowchart of SPS++ Algorithm }
\label{adaptive_sps_algorithm_fig}
\centering
\vspace{-0.2 in}
\end{figure}

\subsection{SPS++ Algorithm Description}
This subsection discusses in detail the proposed SPS++ algorithm. As shown in Fig. \ref{adaptive_sps_algorithm_fig}, SPS++ makes significant enhancements to the conventional SPS algorithm. The dotted boxes represent the new or modified steps in SPS++ (not present in conventional SPS), whereas the solid boxes represent the steps borrowed from conventional SPS. 


  
\begin{itemize}
    
    \item \textbf{RRI Initialization and Sensing (Steps 1-2):} Similar to SPS, SPS++ continuously monitors the previous $N_{sensing}$ subframes by measuring RSRP and S-RSSI and stores the sensing measurements for the sensing window, i.e., $N_{sensing}$ subframes (Step 1). In Step 2, SPS++ initializes the estimated $\widehat{\textnormal{RRI}}$ to $\textnormal{RRI}_{min}$. SPS++ also initializes the RSRP threshold ($P_{th}$) to a minimum value $P_{min}$. Unlike SPS, the selection window is not initialized before starting the resource selection (see Step 2 in Fig. \ref{SPS_algorithm_fig}), and is varied as available resources are identified and the estimated $\widehat{\textnormal{RRI}}$ is updated.

    \item \textbf{Identifying available resources under the chosen RRI (Steps 3-10):}  Each vehicle utilizes $N_{sensing}$ subframes (obtained in Step 2) for identifying the available subframes and subsequently, selecting the minimum RRI possible in between transmission while ensuring that there remain resources (or subframes) for other vehicles. 
    \begin{enumerate}
        \item Like SPS, SPS++ sets and updates the $P_{th}$(See Step 3). However, in Step 4, SPS++ updates the estimated $\widehat{\textnormal{RRI}}$ and initializes the selection window with $T_1=1$ and $T_2=\widehat{\textnormal{RRI}}$. $T_1$ is fixed to 1 to maximize subframe resources. 
        
        \item Similar to SPS, each vehicle populates set $S_A$ with all subframes in the selection window and $S_B$ as a empty set (See Step 5). The candidate subframe exclusion criteria is also borrowed from SPS, except that, $\widehat{RRI}$ is an adjustable parameter in SPS++.
        

        
        \item If the remaining subframes in $S_A$ is less than $20\%$ of total available subframes (Step 7), then, SPS++, unlike SPS, first checks whether the $\widehat{\textnormal{RRI}} < \textnormal{RRI}_{max}$ (Step 8). If yes, $\widehat{\textnormal{RRI}}$ is increased by $\Delta$ as shown in Step 9, and Steps 4 - 7 are repeated. Once $\widehat{\textnormal{RRI}}$ has reached $\textnormal{RRI}_{max}$, then $P_{th}$ is increased by 3 dB in Step 10, and Steps 3 - 7 are repeated.
        
        
    \end{enumerate}

    \item \textbf{Resource Selection} (Steps 11 - 12) and \textbf{Resource Reselection} (Step 13) are similar to SPS. However, in case of SPS++, we choose a reselection counter (RC) value such that the resource reservation is restricted to $0.5$ s, irrespective of chosen $\widehat{\textnormal{RRI}}$. Once RC is zero, unlike SPS (see Steps 12 - 13 in SPS flowchart), SPS++ does not allow reservation of subframe resources. Both these modifications are to ensure that SPS++ allows each vehicle to adjust its RRI at every $0.5$ s and account for changing vehicle traffic conditions.

    
\end{itemize}

\section{Simulation Description and Results}
\label{Section- Simulation Description and Results}
\vspace{-0.05in}

\par This section describes the simulation setting, followed by experimental results for the proposed SPS++ compared against conventional SPS protocol.


\subsection{Simulation Setting}





Both SPS and SPS++ protocols are implemented on top of a modified system-level network simulator (ns-3) which supports C-V2X mode-4 communications. NIST originally implemented Mode 1 and 2 of the LTE sidelink in Network simulator (ns-3)~\cite{NIST_D2D}, which was extended to support Mode-3 and Mode-4 C-V2X Communication by Nabil et al. \cite{Nabil_SPS}.

The highway mobility model assumes vehicles to be moving in a six lane highway road, with three lanes in each direction. While existing work \cite{magazine_paper}, \cite{Nabil_SPS} and \cite{3gpp_V2V} largely assume a constant velocity model, we consider a more realistic vehicle velocity model where the velocity of a certain vehicle follows a Gaussian distribution centered around $v_{avg}$ (in forward direction, i.e., lane 1, 2 and 3) and $-v_{avg}$ (in reverse direction, i.e., lane 4, 5, and 6). In our experiments, the mean $v_{avg}$ is set to $19.44$ m/s ($70$ km/hr) and the variance is set to $3.0$ m/s. 
\cite{Camp_mobility_2002} supports the assumption of Gaussian random variables as reliable for modelling highway traffic speeds. When each vehicle reaches the end of the highway segment, a warp is applied that moves the vehicle immediately to the other end of road segment, and the vehicle is kept in the same lane. Vehicles with velocities normally distributed around $v_{avg}$ are assigned to lanes 1, 2, and 3 (the 3 bottom most lanes in Fig.1), while vehicles normally distributed around $-v_{avg}$ are assigned to lanes 4, 5, and 6. Each vehicle in the same lane travelled in the same direction and had a maximum deceleration of 4.6 m/s$^2$. A Poisson distribution was used for the initial placement of vehicles along the highway.

\par For extensive analysis, we compare the performance of SPS++ against conventional SPS with three different fixed RRIs: $20$ ms RRI, $50$ ms RRI, and $100$ ms RRI. 
The simulations were run with varying vehicle densities, ranging from $40$ to $160$ vehicles. For the purposes of our simulations we assume a 10 MHz channel, though this can be easily adjusted. The initial positions and velocities were kept the same for the same considered vehicle density scenario across the SPS++ and conventional SPS simulations. 
Each scenario had a simulation time of $8$ seconds, and results were averaged across 10 trials. Table \ref{TABLE_1} summarizes the key simulation parameters for both SPS++ and conventional SPS with fixed RRIs. 
\par




\begin{table}
\caption{Simulation Parameters}
\label{TABLE_1}
\vspace{-0.1in}
\centering
\begin{tabular}{ |c|c|}
\hline
Parameter & Value\\
\hline
Vehicle density & \{20-80\} veh/km\\\hline
Road Length/Number of lanes/Lane width & 2000 m/ 6 lanes/ 4 m\\\hline
Simulation time & 8 seconds\\\hline
Transmission Power & 23 dBm\\\hline
Transmission and Sensing Range & 300 meters\\\hline
Distribution of vehicle speeds &  $N$(19.44 m/s, 3 m/s)\\\hline
Deceleration $a$ & 4.6 m/$s^2$\\\hline
$t_{react}$ & 1 second\\\hline
SPS++ RRI$_{max}$ & 100 ms\\\hline
SPS++ RRI$_{min}$ & 20 ms\\\hline
Packet Size & 190 B\\\hline
MCS Index & 7\\\hline
Propagation Model & Winner+ B1 Model\\\hline
Number of trials & 10\\\hline
RSRP threshold, $P_{th}$ & -110 dBm\\\hline
\end{tabular}
\vspace{-0.2in}
\end{table}

 \begin{figure*}[!h]
 \vspace{-0.05in}
 \centering
 \subfigure[\label{fig:RRI_time}]{
 \epsfig{figure=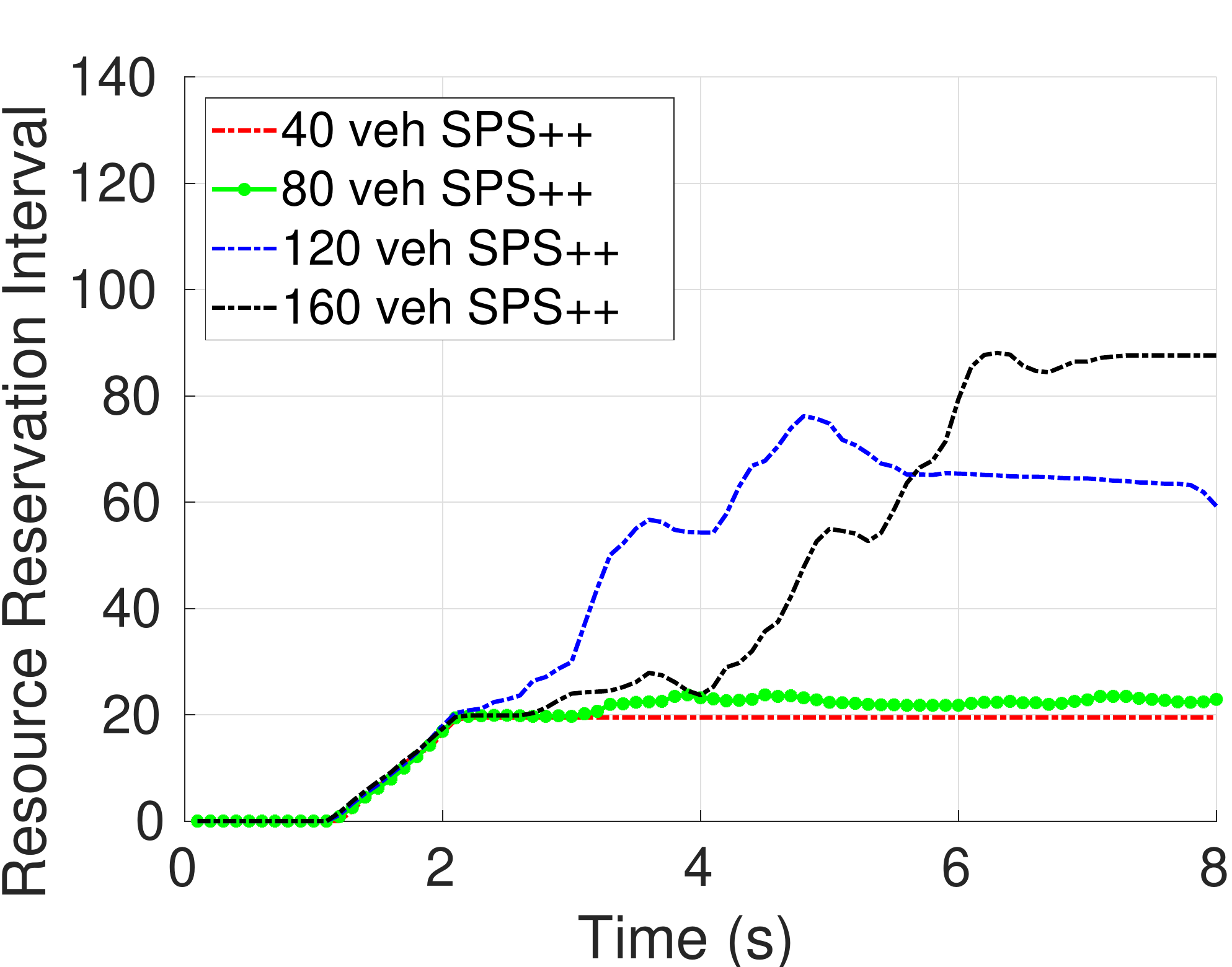,width=2.0 in,  keepaspectratio}}
 \subfigure[\label{Tracking_error}]{
 \epsfig{figure=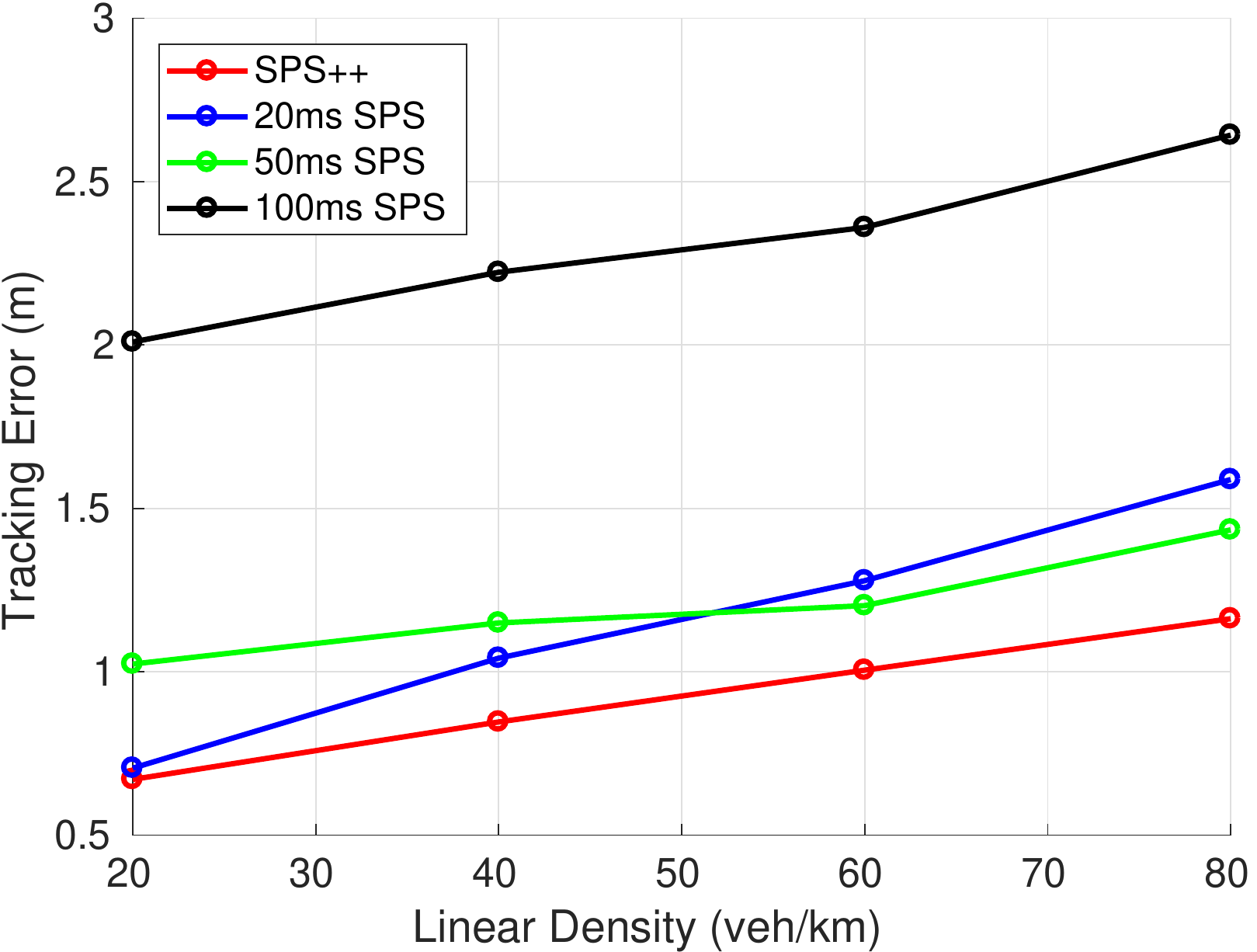,width=2.0 in,  keepaspectratio}}
 \subfigure[\label{Collision_risk}]{
 \epsfig{figure=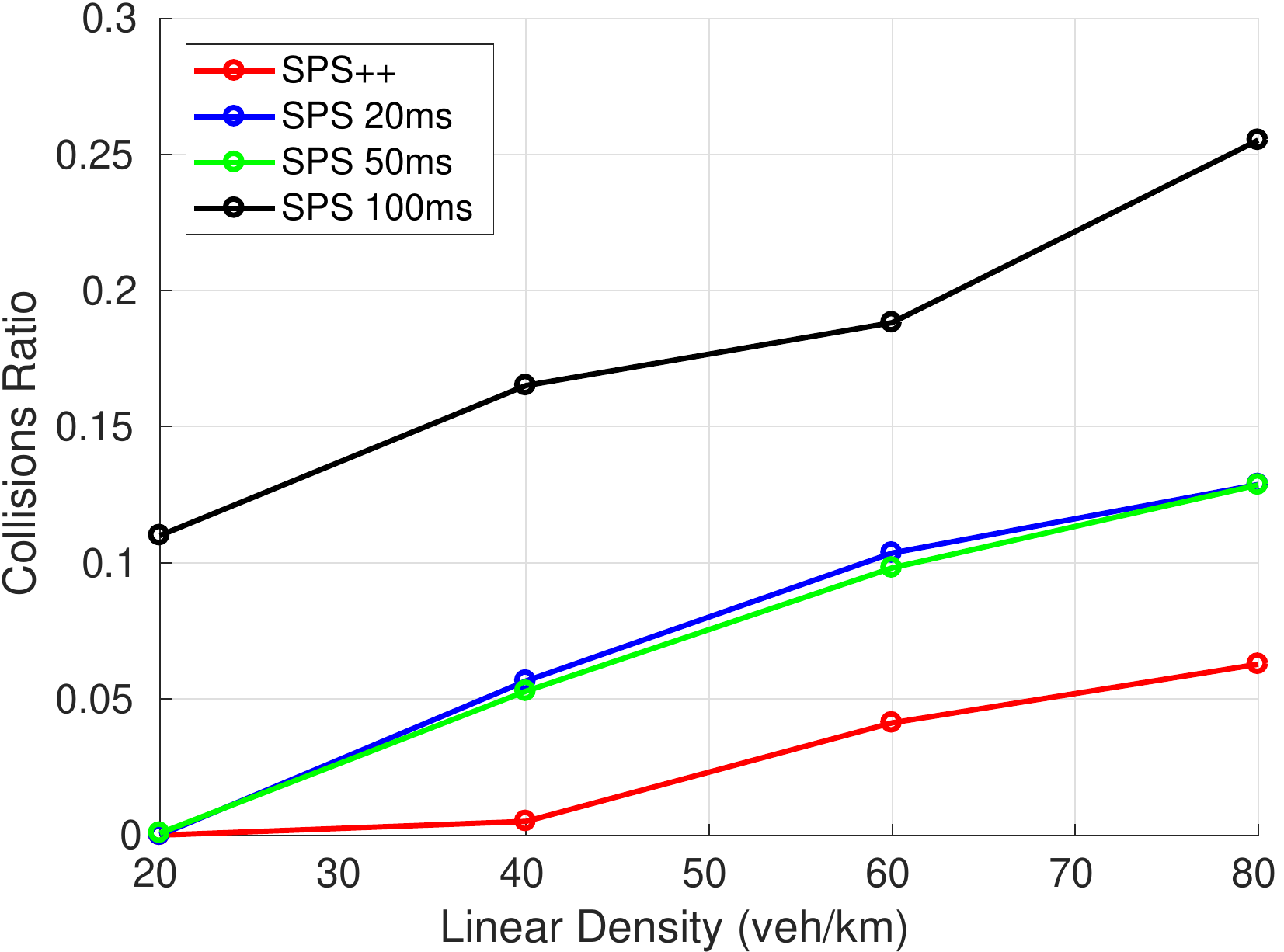,width=2.0 in,  keepaspectratio}}
 \vspace{-0.1in}
 \caption{(a) RRI chosen by SPS++ over time and varying vehicle densities: (b) Average high risk tracking error, and (c) Collision Risk}
 \vspace{-0.15in}
 \end{figure*}

\subsection{Performance Metrics}


\par We use the following two performance metrics for the evaluation of SPS++ against conventional SPS. 

\begin{itemize}
     \item \textbf{Tracking error ($\mathbf{e_{track}}$)}- $e_{track}$ is the difference between transmitting vehicle $u$'s actual location and $u$'s location obtained from the most recent BSM received from $u$ at receiver vehicle $v$. (see Section \ref{Subsection:Tracking Error}).
    \item \textbf{Collision Risk Ratio} - 
    It measures the overall on-road safety performance of Mode-4 C-V2X networks, and is defined as the ratio of number of collision risky instances (computed using Eq. \ref{Collision_risk_eq_2}) to that of both non-risky and risky instances between each pair of vehicles.
    
    
    \item  \textbf{Packet Delivery Ratio (PDR)} - The probability that all vehicles within the range of the transmitting vehicle receives the transmitted packet. The PDR is calculated as PDR$_u=\frac{PR_u}{PD_u}$, where $PD_u$ is the number of BSMs transmitted by vehicle $u$ and $PR_u$ is the number of BSMs sent by vehicle $u$ received by neighboring vehicles. 

\end{itemize}

\subsection{Experimental Results}

Before we discuss the comparative analysis of SPS++ and SPS in terms of aforestated performance metrics, we first briefly discuss how SPS++ is able to adjust the RRI at each vehicle given varying vehicle densities and over time.

Fig. \ref{fig:RRI_time} shows how SPS++ chooses, on average, the RRI across all vehicles for each considered vehicle density scenario (i.e., $40$, $80$, $120$ and $160$) under varying simulation time.  We observe over 10 trials that the average chosen RRI increases over simulation time as vehicles enter the simulation for each considered vehicle density.
After $2$ seconds, the RRI chosen for each vehicle setting converges to a certain unique RRI value. The chosen RRI in case of $40$ and $80$ vehicle density (i.e., sparse) is almost four to five times than that of $120$ or $160$ (i.e., dense) vehicle setting, thanks to the ability of SPS++ to adjust RRI in real-time and adapt to the varying C-V2X environment. Vehicles also take a longer time in dense vehicle settings ($5$ and $6$ seconds for 120 and 160 vehicles respectively) to converge to a chosen RRI value, likely because it takes longer for all vehicles to enter the simulation.   



\begin{figure}[!h]
\subfigure[$40$ vehicles \label{fig:Distributions-40 veh}]{
\epsfig{figure=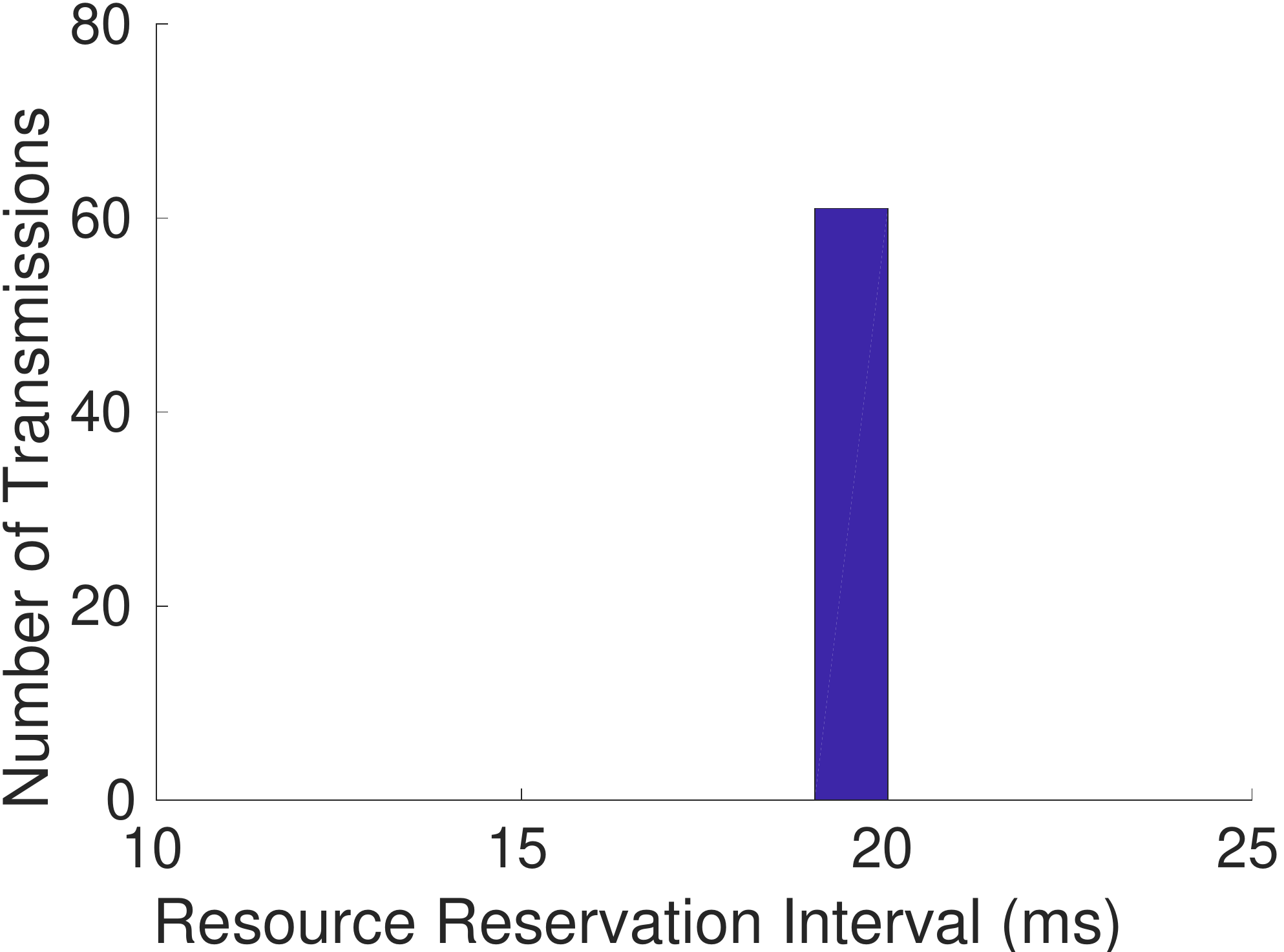,width=1.62 in,  keepaspectratio}}
\subfigure[$80$ vehicles \label{fig:Distributions-80 veh}]{
\epsfig{figure=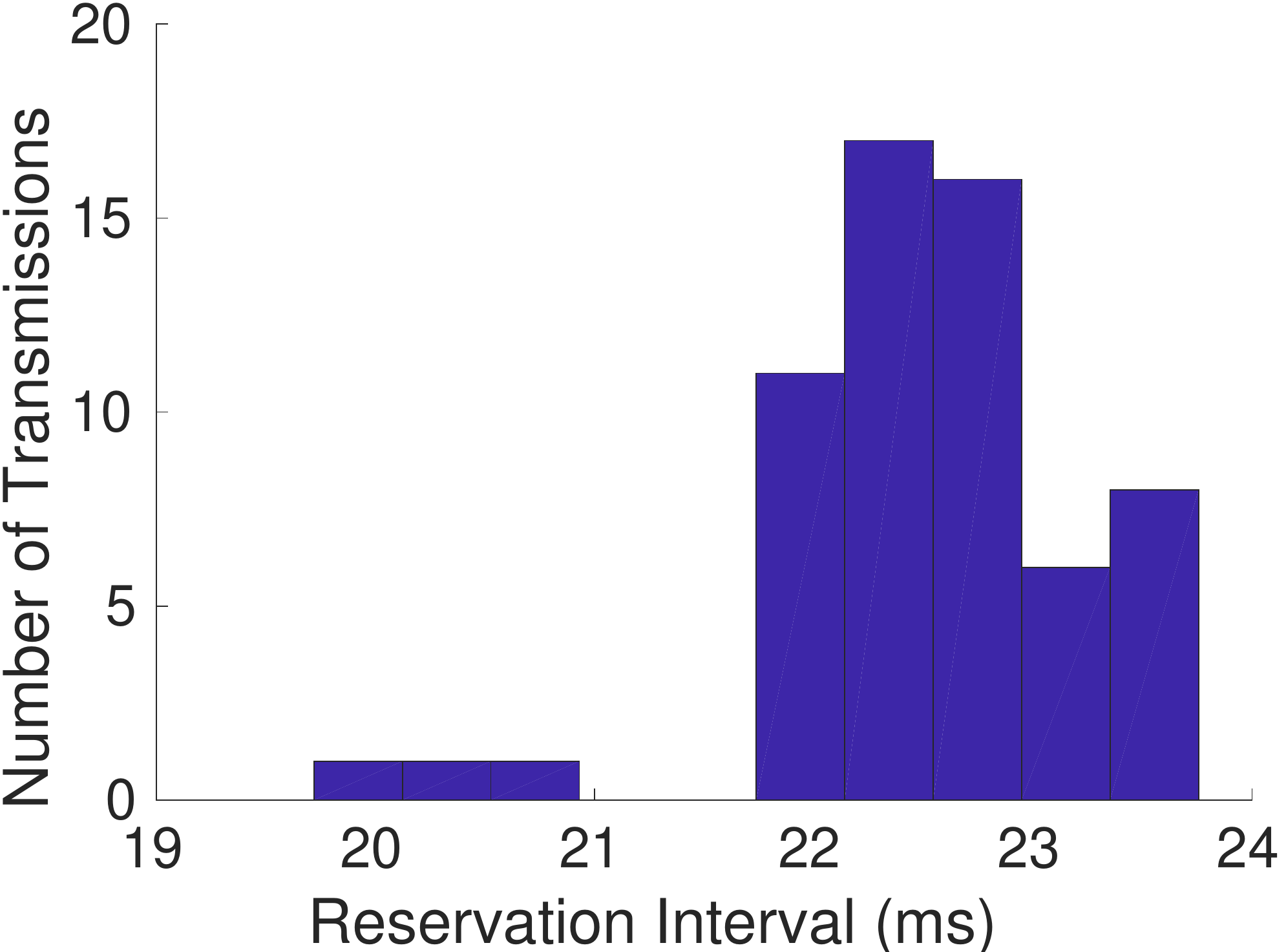,width=1.62 in,  keepaspectratio}}
\subfigure[$120$ vehicles \label{fig:Distributions-120 veh}]{
\epsfig{figure=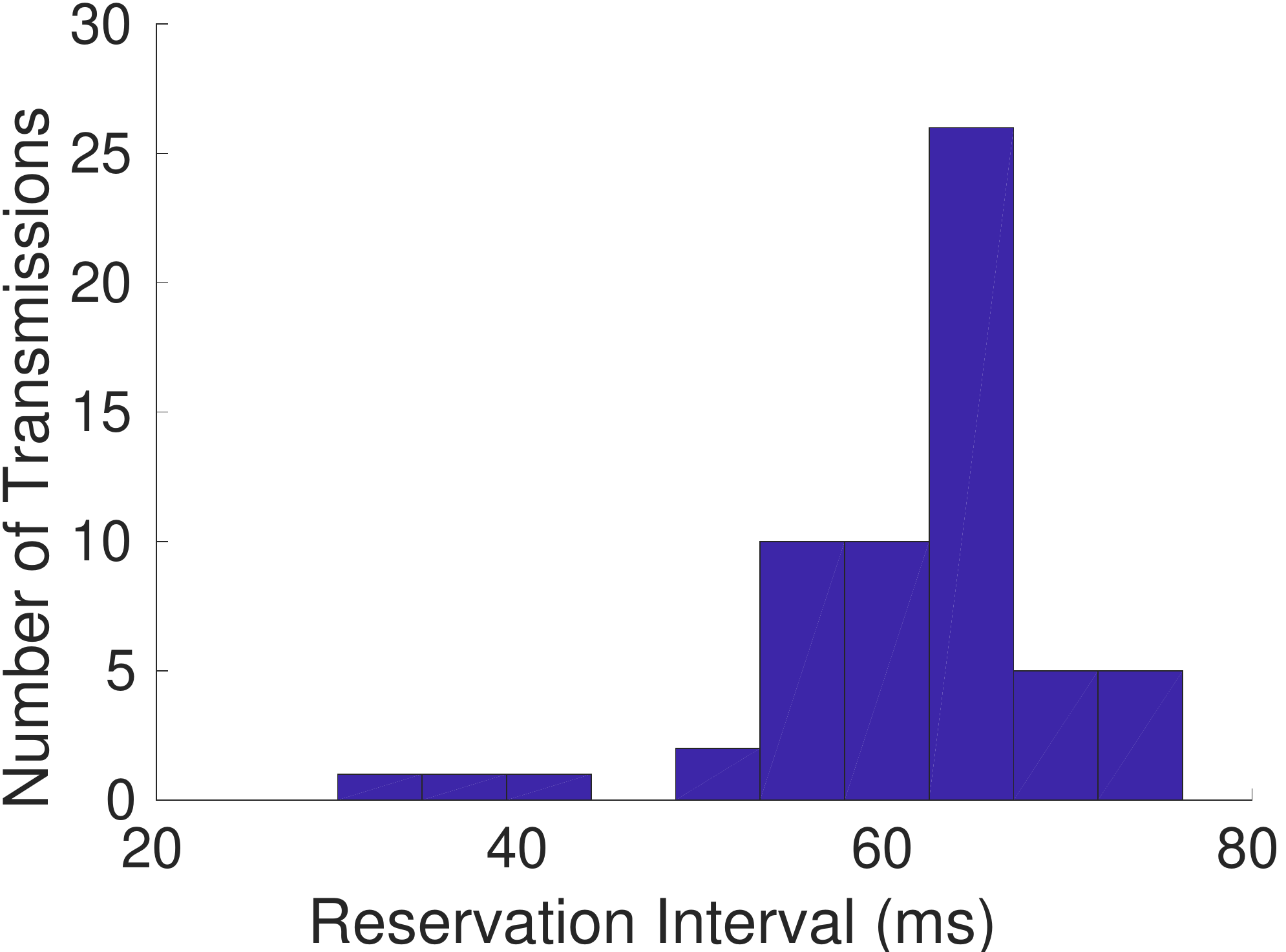,width=1.62 in,  keepaspectratio}}
\subfigure[$160$ vehicles \label{fig:Distributions-160 veh}]{
\epsfig{figure=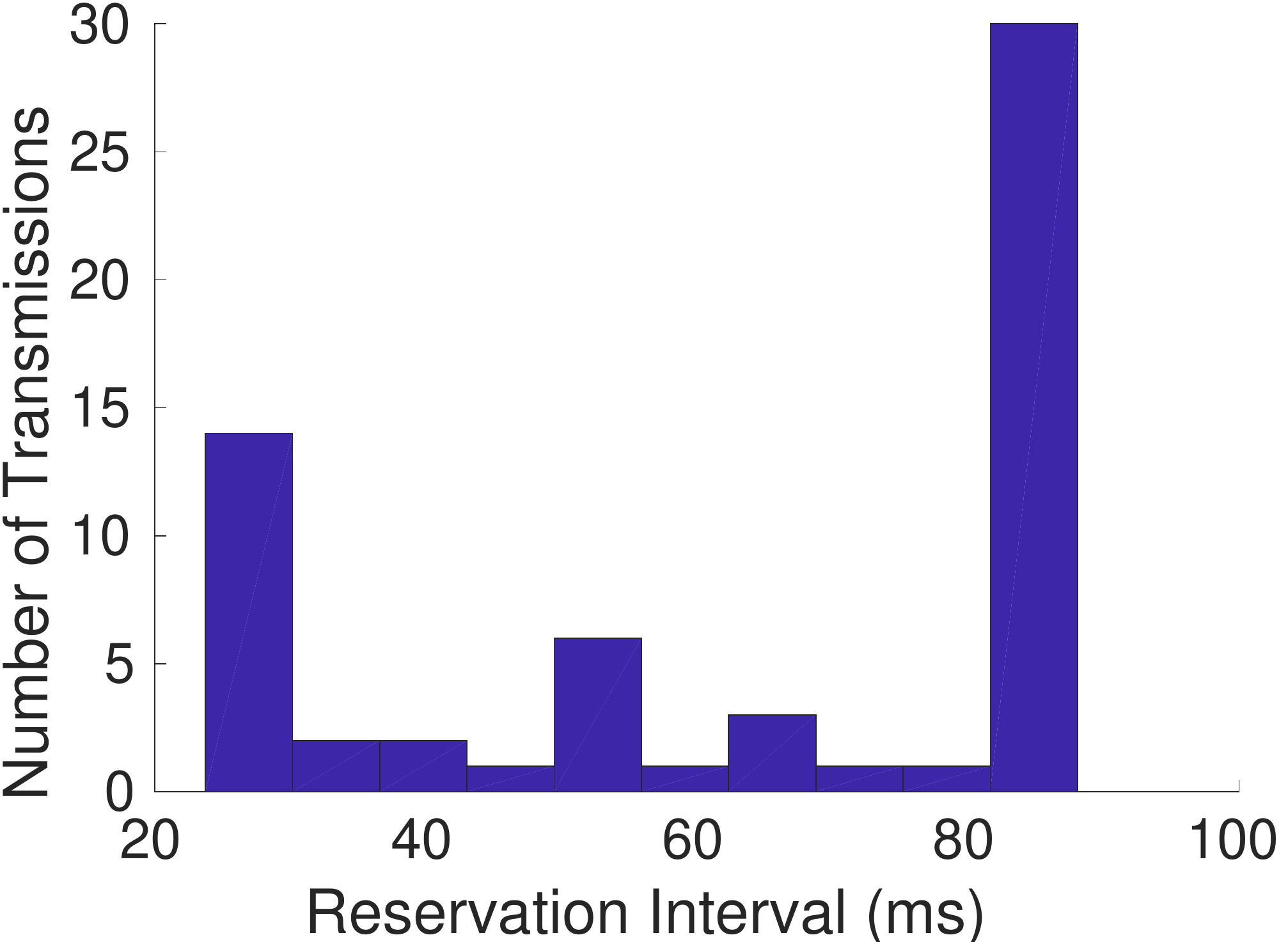,width=1.62 in,  keepaspectratio}}
\vspace{-0.09in}
\caption{ Average RRI distribution across different vehicle densities}\vspace{-0.18in}
\end{figure}

Figs. \ref{fig:Distributions-40 veh}-\ref{fig:Distributions-160 veh} show average SPS++ RRI distributions over the last second of the simulation~\footnote{The reason we use the last second is because by this time, all vehicles have entered the simulation, and chosen an suitable RRI. See Fig. \ref{fig:RRI_time}} and show a large RRI variance in the $80$, $120$ and $160$ vehicle density simulations. Not that the chosen RRI is unique
This is due to the varying vehicle traffic densities, and vehicles choosing different RRIs depending on the number of vehicles in their vicinity. The average RRI of $40$ vehicles, on the other hand, converges to the $RRI_{min}$ of $20$ ms. This is because there are sufficient channel resources to support $20$ ms RRI for each vehicle in case of $40$ vehicles, irrespective of varying neighborhood sizes. Notice from Figs. \ref{fig:Distributions-120 veh} and \ref{fig:Distributions-160 veh} that while the RRI distribution is skewed to RRI$_{max}$, there are a number of vehicles still transmitting at lower RRIs. 

\textbf{Tracking Error and Collision Risk Analysis.} The tracking error (Fig. \ref{Tracking_error}) and the collision risk (Fig. \ref{Collision_risk}) are both used to compare the effectiveness of SPS++ to $20$ ms, $50$ ms, and $100$ ms fixed RRI SPS, in terms of improved road safety performance of C-V2X networks. Fig. \ref{Tracking_error} shows that the tracking error increases as the vehicle density increases, and $20$ ms RRI SPS and $50$ ms RRI SPS tracking error alternate as the best performing fixed RRI SPS. At higher vehicle density scenarios, $50$ ms RRI performs better in terms of tracking error, likely because the $20$ ms RRI leads to increased packet losses at higher densities. Notice that the tracking error for SPS++ is significantly lower than the tracking error associated with SPS (either with RRI $20$ ms, $50$ ms or $100$ ms) across all vehicle densities. On average, SPS++ outperforms $50$ ms RRI SPS (the best performing fixed RRI) by almost $17\%$ and $23\%$ repectively at $60$ and $80$ veh/km. This indicates that SPS++ protocol is able to choose a suitable RRI at each vehicle that minimizes packet losses and enable up-to-date BSM sharing between a certain vehicle and its neighboring nodes. The tracking error improvements found with SPS++ expectedly to extend to the collision risk. Fig. \ref{Collision_risk} shows a significant decrease in high risk scenarios at $60$ and $80$ veh/km ($55.6$ \% and $51.20$ \% respectively) as compared to $50$ ms RRI SPS, which is the best performing fixed RRI scheme. 

These results shows that the lower RRI chosen is able to help vehicles in both low and high density situations. Observe that even at highly dense situations ($80$ vehicles/km), SPS++ outperforms the $100$ ms SPS protocol, although the average RRI chosen for SPS++ was $100$ ms. It is suspected that a small number of vehicles could be using smaller RRIs during the course of the simulation which help reduces the number of high risk situations. 



 \begin{figure}[htbp]
 \vspace{-0.14in}
 \subfigure[\label{fig:40_veh_PDR}]{
 \epsfig{figure=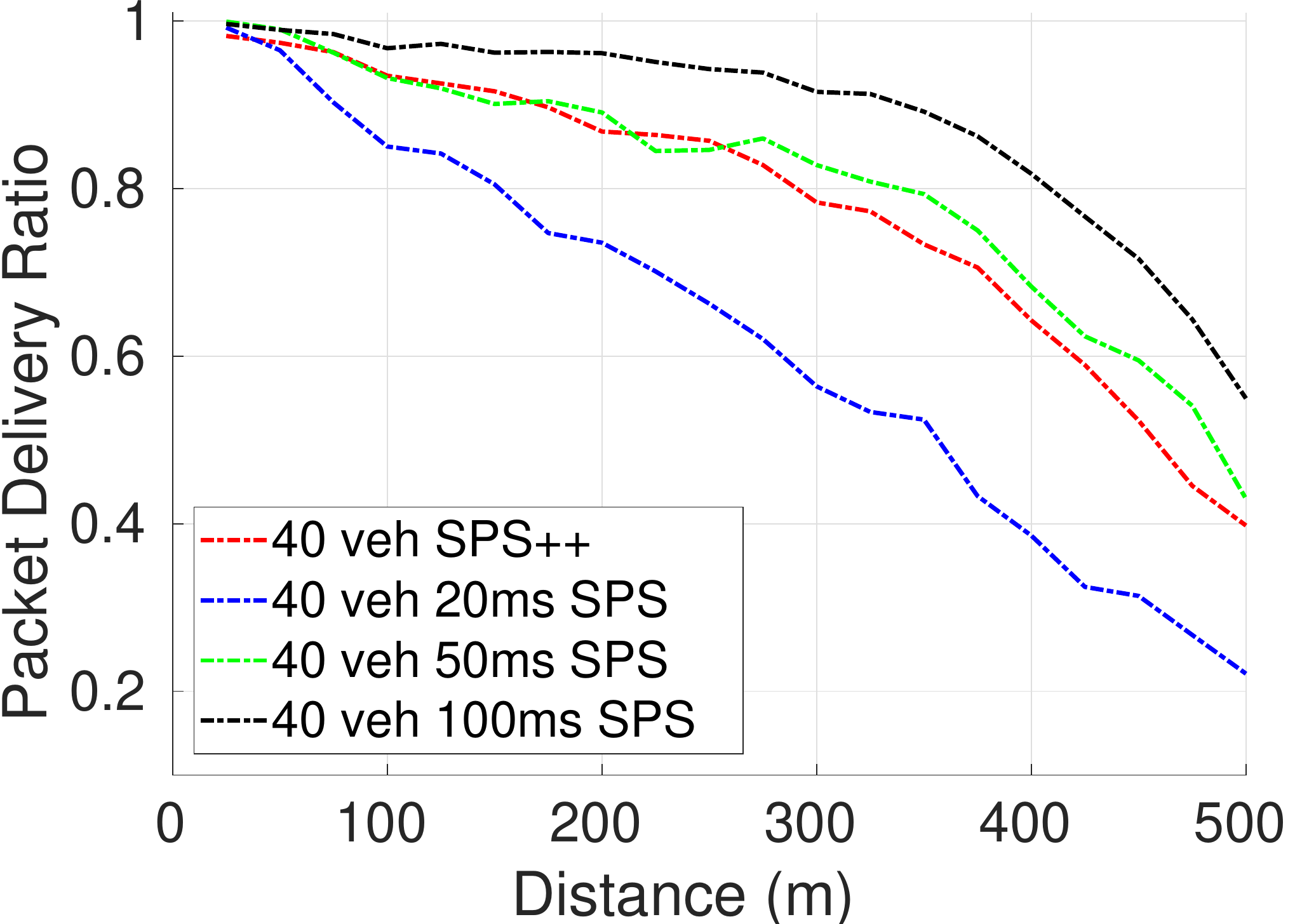,width=1.6 in,  keepaspectratio}}
 \subfigure[\label{fig:80_veh_PDR}]{
 \epsfig{figure=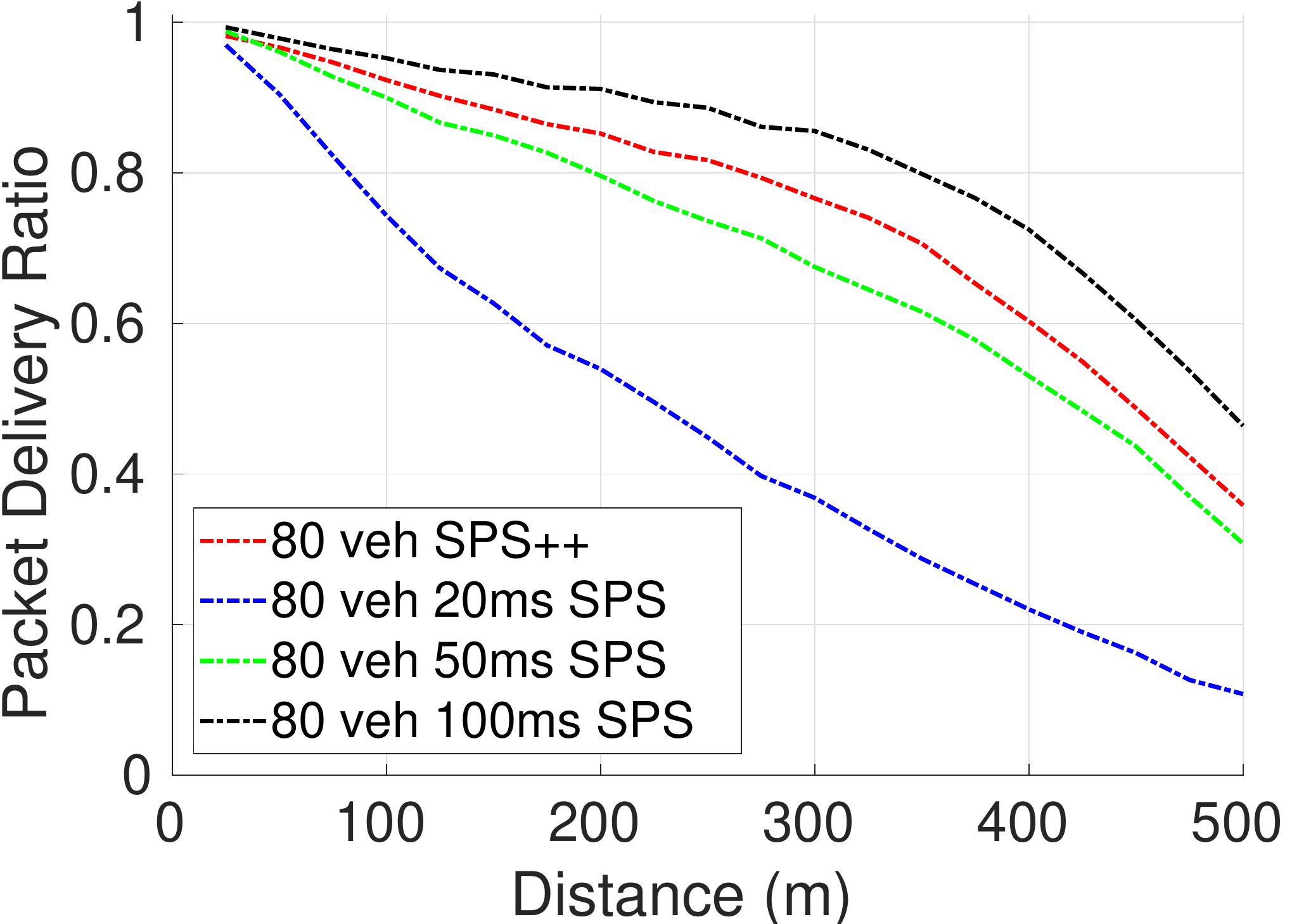,width=1.6 in,  keepaspectratio}}
  \subfigure[\label{fig:120_veh_PDR}]{
 \epsfig{figure=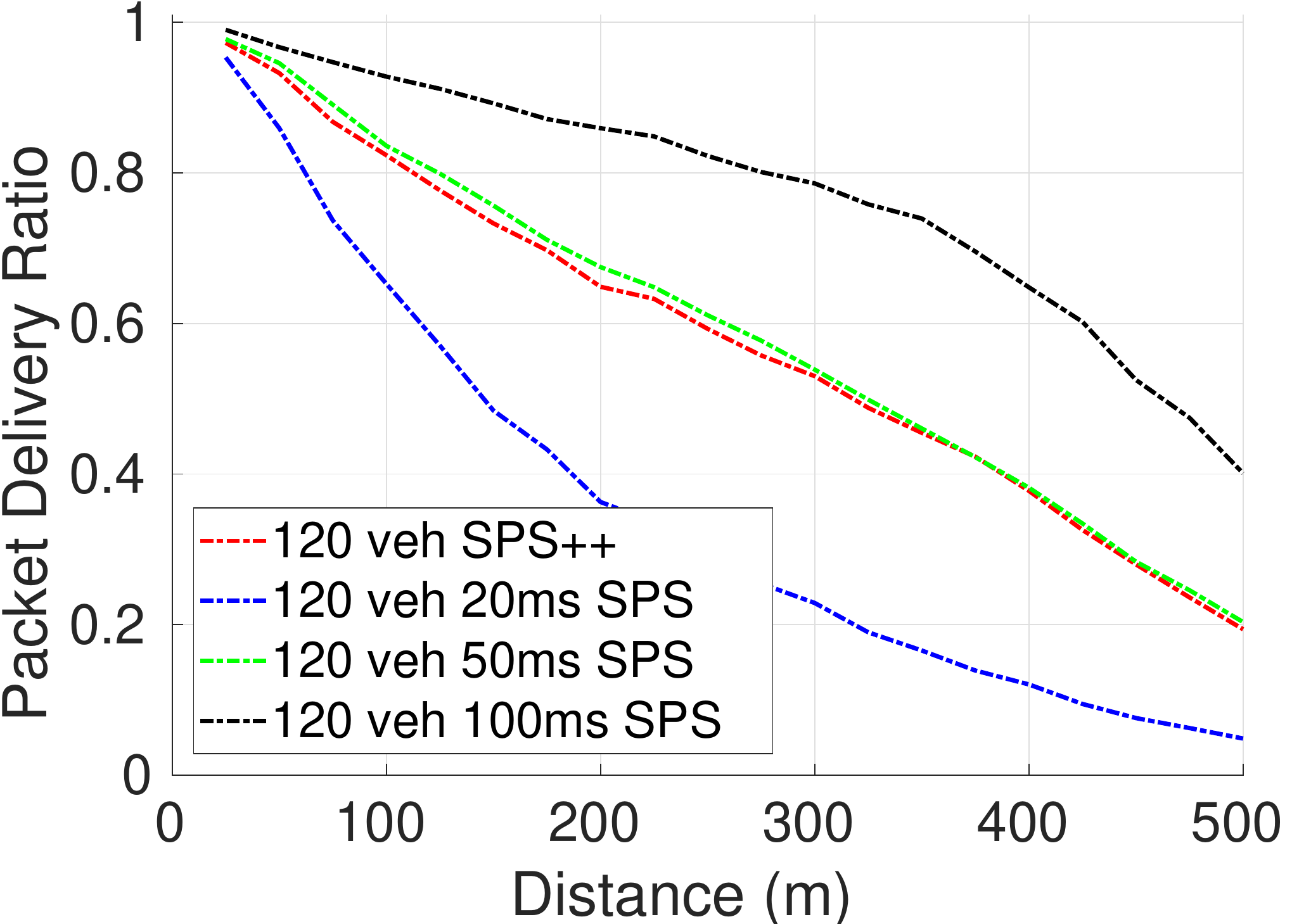,width=1.6 in,  keepaspectratio}}
 \subfigure[\label{fig:160_veh_PDR}]{
 \epsfig{figure=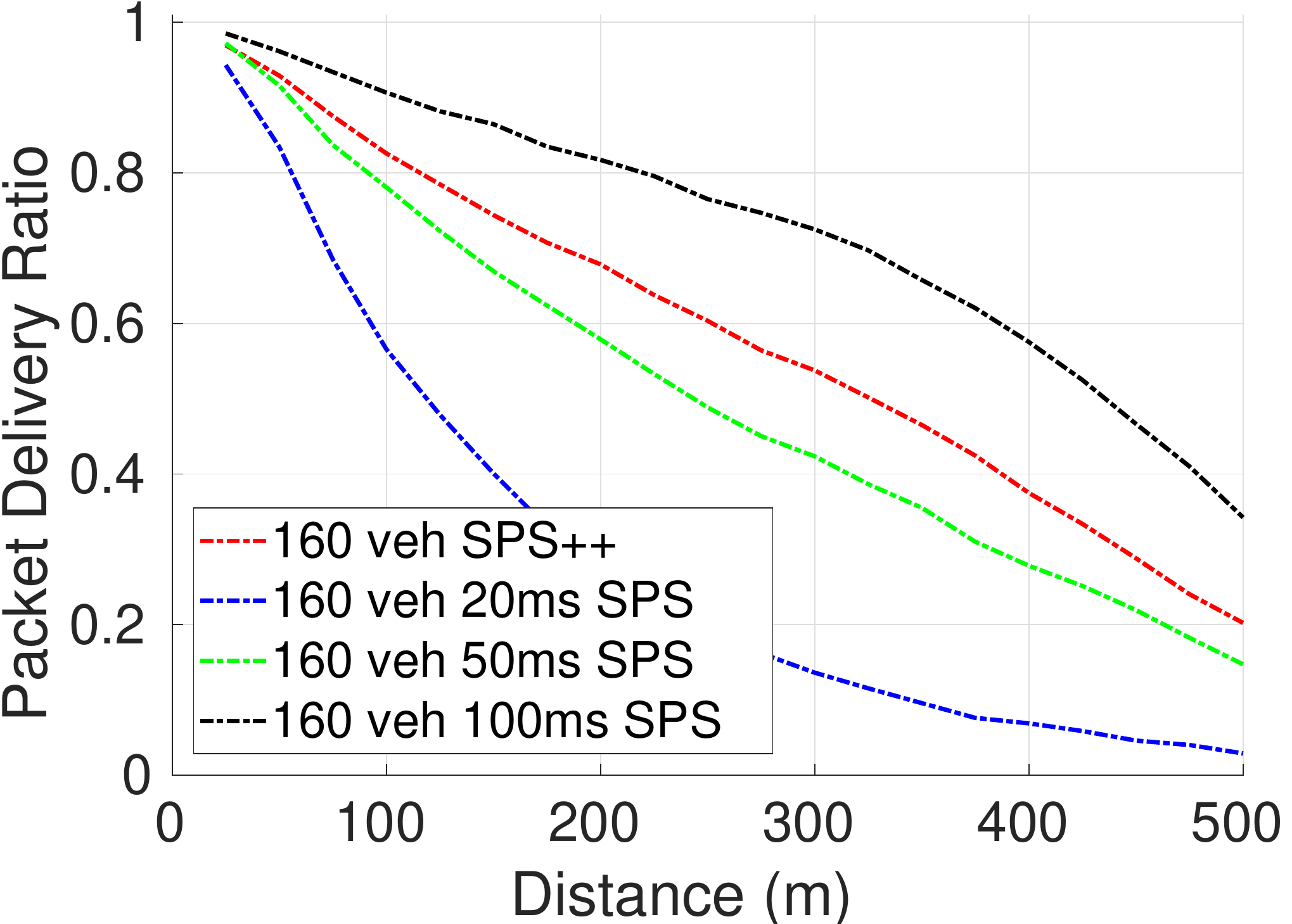,width=1.6 in,  keepaspectratio}}
  \vspace{-0.09in}
 \caption{ (a) PDR for 40 vehicles (b) PDR for 80 vehicles (c) PDR for 120 vehicles and (d) PDR for 160 vehicles} \label{fig:PDR_results}
 \vspace{-0.10in}
 \end{figure}
\textbf{Packet Delivery Ratio Analysis.} Fig. \ref{fig:PDR_results} shows that SPS++ performs significantly better in terms of PDR when compared to that of SPS with RRIs $20$ ms and $50$ ms, under all considered vehicle density scenarios. 
Among the SPS with different RRIs, $100$ ms SPS PDR performs the best of all three fixed RRIs. Larger RRIs led to a better PDR, but also yield a large tracking error and collision risk (See Figs \ref{Tracking_error} and \ref{Collision_risk}). Also notice that other than $100$ ms RRI SPS, none of the $20$ ms RRI, $50$ ms RRI, or SPS++ protocols have PDRs near 100\%, although at low densities there should be enough subframes for all vehicles. This reduced PDR is likely an effect of the hidden terminal problem, where two transmitters might be out of the sensing range of each other, even if they are both transmitting to the same destination. 
The likelihood of a hidden terminal collision increases with smaller RRIs, as there are fewer subframes that vehicles can use. However, in SPS++ the PDR remains the same across vehicular densities, and outperforms the $50$ ms fixed RRI PDR at higher densities. This is likely because of the larger RRI chosen by SPS++ in the dense scenarios, which likely leads to less congestion and fewer packet collisions.

\textbf{Discussion.} Unlike conventional SPS protocol, the proposed adaptive SPS++ is successfully able to adapt to the considered time-varying C-V2X scenarios and 
allows each vehicle to dynamically choose the best RRI for the efficient and reliable BSM sharing with neighboring vehicles 
This greatly reduces the tracking error between each pair of vehicles, which in turn significantly enhances the road safety of C-V2X networks. Thus, SPS++ is a significant step forward for enabling the next-generation of C-V2X mode-4 based connected vehicles and intelligent transportation systems.

\section{Conclusion} \label{Section-Conclusion}

In this work, we proposed an adaptive, sensing-based semi-persistent scheduling protocol, named, SPS++ for improved on-road safety performance of decentralized V2X networks. Specifically, SPS++, unlike conventional SPS, allows each vehicle to dynamically adjust RRIs based on the availability of channel resources and select suitable transmission opportunities for timely BSM transmissions at adjusted RRIs, while accounting for various vehicle traffic scenarios. Our extensive experiments based on Mode-4 C-V2X standard implemented using ns-3 simulator demonstrated that SPS++ protocol significantly outperformed conventional SPS in terms of improved on-road safety performance in all considered C-V2X scenarios. In the future, we will explore designing reinforcement learning (RL) and deep RL, based BSM scheduling protocol that learns vehicle traffic patterns and other contextual factors over time, and selects suitable RRIs.


\bibliographystyle{IEEEtran}
%
\bibliography{main}

\end{document}



   


 